\documentclass[%
 aip,
 amsmath,amssymb,
reprint, onecolumn %%%%% Comment onecolumn to override one column format %%%%%%%
]{revtex4-1}

\usepackage{graphicx}% Include figure files
\usepackage{dcolumn}% Align table columns on decimal point
\usepackage{bm}% bold math
\usepackage[utf8]{inputenc}
\usepackage[T1]{fontenc}
\usepackage{mathptmx}
\usepackage{etoolbox}
\usepackage{CJK}
\usepackage[mathlines]{lineno}% Enable numbering of text and display math
\usepackage{amsmath}
\usepackage{amssymb}
\usepackage{amsfonts}
\usepackage{mathrsfs}
\usepackage{color}
\usepackage{enumitem}
\usepackage[utf8]{inputenc}
\usepackage[T1]{fontenc}
\usepackage{etoolbox}
\usepackage{subfigure}
\newcommand\etal{{\it et~al.\/}}
\def\BPS{\begin{psfrags}}
\def\EPS{\end{psfrags}}
\def\BC{\begin{center}}
\def\EC{\end{center}}
\def\BI{\begin{itemize}}
\def\EI{\end{itemize}}
\def\BEN{\begin{enumerate}}
\def\EEN{\end{enumerate}}
\def\BENA{\begin{enumerate}[label=(\alph*)]}
\def\EENA{\end{enumerate}}
\def\BENR{\begin{enumerate}[label=(\roman*)]}
\def\EENR{\end{enumerate}}
\def\BE{\begin{equation}}
\def\EE{\end{equation}}
\def\BA{\begin{eqnarray}}
\def\EA{\end{eqnarray}}
\def\BSE{\begin{subequations}}       
\def\ESE{\end{subequations}}
\def\BAN{\begin{eqnarray*}}
\def\EAN{\end{eqnarray*}}
\def\NNM{\nonumber}
\def\P{\partial}
\def\i{{\rm i}}

\makeatletter
\def\@email#1#2{%
 \endgroup
 \patchcmd{\titleblock@produce}
  {\frontmatter@RRAPformat}
  {\frontmatter@RRAPformat{\produce@RRAP{*#1\href{mailto:#2}{#2}}}\frontmatter@RRAPformat}
  {}{}
}%
\makeatother
\begin{document}
\begin{CJK*}{UTF8}{bsmi}
\preprint{AIP/123-QED}

\title[Stability Analysis of Supersonic Compressible Flows]{Nonlinear Stability and Dynamics of Supersonic 
Compressible Flows}
% \title[Stability Analysis of Supersonic Compressible Flows]{Stability Analysis of Compressible Supersonic Flows: 
% Analytical Modeling and Instability Regimes}
% \title[Stability Analysis of Supersonic Compressible Flows]{Stability Analysis of Compressible Supersonic Flows: 
% An Analytical Exploration of Instability Regimes under Mach Number Variations}
% \title[Weakly Non-Linear Stability Analysis of Supersonic Compressible Flows]{Weakly Non-Linear Stability and 
% Bifurcation Analysis of Compressible Supersonic Flows with Mach Number Variations}

\author{Symphony Chakraborty (辛芬妮)}
% \email{symphony.chakraborty@gmail.com}
\affiliation{ 
Institute of Astronomy and Astrophysics, Academia Sinica, Taipei 106216, Taiwan, (R.O.C.)
%\\This line break forced with \textbackslash\textbackslash
}%
\author{Hsien Shang (尚賢)}%
\email{shang@asiaa.sinica.edu.tw}
\affiliation{ 
Institute of Astronomy and Astrophysics, Academia Sinica, Taipei 106216, Taiwan, (R.O.C.)
%\\This line break forced with \textbackslash\textbackslash
}%

\date{\today}% It is always \today, today,
             %  but any date may be explicitly specified             

\begin{abstract}
The study of shear layer instability in compressible flows is key to understanding phenomena from aerodynamics 
to astrophysical jets. Blumen's seminal paper [``Shear layer instability of an inviscid compressible fluid," 
J. Fluid Mech. {\bf 40}, 769--781 (1970)] established a linear stability framework for inviscid compressible 
shear flows, emphasizing velocity gradients and compressibility effects. However, the nonlinear regime remains 
insufficiently explored. This research extends Blumen's framework by conducting a weakly nonlinear stability 
analysis using the method of multiple scales to derive amplitude equations, such as the Landau-Stuart and 
complex Landau equations. Perturbation variables are expanded in a power series to capture amplitude evolution 
beyond linear theory. Finite boundary conditions are incorporated to enhance physical applicability. The study 
analyzes how compressibility and Mach number influence nonlinear saturation, revealing Mach-dependent bifurcations 
in Kelvin-Helmholtz instability (KHI) with alternating stable and unstable regimes. Phase portraits and trajectories 
illustrate transitions, saturation, and spiral decay, which are relevant to astrophysical shear flows. Bifurcation 
analysis reveals both supercritical and subcritical Hopf behavior in compressible shear flows, underscoring the 
importance of nonlinear effects in the onset and evolution of flow instabilities. Qualitative and quantitative 
results of instability evolution have been shown from nonlinear stability analysis. This work bridges the critical 
gap between linear and fully nonlinear stability analyses by offering a systematic weakly nonlinear framework 
and the nonlinear dynamics for compressible shear layers. It generalizes the earlier linear results and provides 
new predictions about bifurcation behavior and long-time state selection in compressible flows.
\end{abstract}

\maketitle
\end{CJK*}
\section{Introduction\label{sec:intro}}
The stability of parallel shear flows in an inviscid, homogeneous fluid under infinitesimal, two-dimensional, 
non-divergent disturbances has been a prominent topic in scientific research for over a century. While this 
model has limited practical applications, it has significantly contributed to developing mathematical methods 
and a deeper understanding of inertial instability mechanisms. It serves as a valuable foundation for investigating 
more complex and realistic scenarios. The stability of compressible shear flows has long captivated fluid 
dynamicists, beginning with the foundational analytical study by Lees and Lin,\cite{Lees46} who first investigated 
the stability of laminar boundary layers and laid the groundwork for modern compressible flow theory. Chandrasekhar's 
classic monograph synthesized hydrodynamic and hydromagnetic stability analyses, offering deep mathematical insights 
into shear-driven instabilities.\cite{Chandrasekhar61} Extending theoretical boundaries, Eckart\cite{Eckart63} 
analytically generalized Howard's semicircle theorem for adiabatic jets, enriching our understanding of inviscid 
compressible flow behavior. In recent efforts, Chaturvedi \etal\cite{Chaturvedi24} employed the compound 
matrix method to numerically investigate compressible boundary layers, revealing the existence of multiple unstable 
modes beyond Mach $3$. Similarly, Mack\cite{Mack75} provided seminal linear stability results addressing the 
complexities of supersonic boundary-layer transitions. Drazin and Reid\cite{Drazin04} further codified the theory 
of hydrodynamic stability with a comprehensive analytical framework that has influenced numerous subsequent studies. 

Beyond linear theory, El-Hady\cite{Elhady89} explored secondary instabilities in compressible boundary layers through 
a spatial, three-dimensional numerical approach, underscoring the importance of nonlinear effects in high-speed flows. 
Goldstein and Ricco\cite{Goldstein18} analytically examined receptivity-induced non-localized instabilities at 
supersonic Mach numbers, while Faria \etal\cite{Faria15} advanced weakly nonlinear theory in the context of 
self-sustained detonations, highlighting the nonlinear saturation mechanisms. Ladeinde and Wu\cite{Ladeinde02} 
pursued a second-order nonlinear spatial stability analysis for compressible mixing layers, offering numerical 
insights into subharmonic interactions and modal evolution. At hypersonic regimes, Hildebrand \etal\cite{Hildebrand18} 
combined high-fidelity simulations with stability analysis to study shock-wave/boundary-layer interactions near 
Mach $6$, and Karpuzcu \etal\cite{Karpuzcu25} recently investigated linear instabilities in supersonic flows over 
compression corners, illustrating the growing role of numerical simulations in high-speed flow diagnostics.

Blumen's\cite{Blumen70} influential work significantly advanced the theoretical understanding of shear layer stability 
in compressible, inviscid flows. By extending classical incompressible stability theories, he showed that compressibility 
quantified through Mach number, density, and pressure variations introduces stabilizing effects that alter the growth 
of perturbations. Blumen's analysis, based on a hyperbolic-tangent velocity profile and the compressible Euler equations, 
revealed the existence of subsonic neutral mode and demonstrated that at higher Mach numbers, inertial forces weaken 
the interaction between pressure waves and vorticity, thereby reducing instability growth rates. Additionally, strong 
density gradients were shown to either suppress or amplify instabilities depending on their orientation relative to 
velocity gradients. Blumen incorporated new terms in the stability equation, such as pressure gradient effects, to 
account for these compressibility influences. His findings laid a foundational framework for understanding the spectral 
behavior of compressible shear layers, with important implications for high-speed flows in aerodynamics and astrophysics, 
where compressibility plays a crucial role in determining flow stability and transition to turbulence.

Blumen \etal\cite{Blumen75} expanded on earlier studies of compressible shear layer instability by incorporating more 
complex flow profiles and examining the combined effects of compressibility and density stratification. Their analysis 
of the dispersion relation revealed how compressibility introduces stabilizing mechanisms that modify or suppress 
perturbation growth, particularly at high Mach numbers where interactions between vorticity and acoustic wave weaken. 
They also demonstrated how density gradients can either stabilize or destabilize a flow depending on their alignment 
with velocity shear. Through numerical results, they illustrated the intricate balance between shear, stratification, 
and compressibility, providing a more realistic view of high-speed flow behavior in practical applications such as 
aerodynamics and astrophysical systems.

Building on this, Drazin and Davey\cite{Drazin77} focused on asymptotic analyses at high Mach numbers and examined how 
specific flow features, like velocity profiles and stratification, affect stability. They confirmed that compressibility 
has a strong stabilizing influence by damping interactions between pressure waves and vorticity, though they also 
identified conditions under which instabilities could persist despite high-speed stabilization. Their work, along with 
findings by Blumen \etal\cite{Blumen75}, uncovered additional families of unstable modes, including a second and third 
mode of instability, demonstrating that compressible shear flows can remain generically unstable to two-dimensional 
perturbations even under strong compressibility, broadening the known spectrum of instability behavior in such systems.

While these studies significantly enriched the linear stability landscape, they left open important questions regarding 
the nonlinear evolution of instabilities, particularly in compressible regimes. In linear analyses, modal growth rates 
and eigenfunction structures are informative; however, they do not capture the amplitude saturation, mode interaction, or 
nonlinear feedback mechanisms that may govern the long-time behavior of disturbances. The transition from linear 
instability to finite-amplitude states remains inadequately understood within the compressible framework. In this 
context, the present study offers a novel contribution by extending the classical linear theory into the weakly 
nonlinear regime through a Landau-Stuart amplitude expansion. By employing a multiple scales perturbations approach, 
we derive and analyze the amplitude evolution equation near the marginal stability threshold. This allows us to 
characterize the bifurcation structure, identify critical Mach numbers corresponding to supercritical or subcritical 
transitions, and uncover secondary saturation mechanisms that cannot be captured by linear theory alone. Unlike the 
previous studies,\cite{Blumen70,Blumen75,Drazin77} which focused purely on eigenvalue spectra and mode multiplicity, 
our analysis captures the nonlinear saturation of unstable modes and provides insight into the post-instability 
dynamics. Furthermore, we compute the Landau coefficient explicitly as a function of compressibility (Mach number) 
and investigate its role in determining the nature of the bifurcation. Our results include detailed phase portraits, 
three-dimensional trajectories, and time evolution of the amplitude, offering a dynamics systems perspective on shear 
layer instability. In addition, our analysis revisits the boundary conditions employed in the classical linear theory, 
adopting a novel yet physically consistent boundary framework that better captures the nonlinear feedback and flow 
confinement effects. These modifications yield richer dynamics and enable more accurate modeling of realistic physical 
systems.

The structure of this paper is organized as follows: In Sec. \ref{sec:maths}, we present the mathematical formulation 
of the problem, outlining the governing equations for compressible shear flow and the assumptions underlying the model. 
Section \ref{sec:wnsa} details the weakly nonlinear stability analysis, including the derivation of the amplitude 
equations and the exploration of nonlinear effects near the instability threshold. In Sec. \ref{sec:bifurcation}, the 
nonlinear dynamics and bifurcation characteristics of the compressible Kelvin-Helmholtz instability (KHI) are 
examined through bifurcation analysis, revealing the influence of Mach number and density stratification on flow 
behavior. Finally, Sec. \ref{sec:conslusions} offers the conclusions, summarizing the key findings and discussing their 
implications for high-speed shear flow stability.

\section{Mathematical Formulation\label{sec:maths}}
As we are dealing with compressible fluid, we have
\BA
\label{floweq}
p^* &=& f(\rho^*, S)\,,
\EA
where $p^*$, $\rho^*$, and $S$ denote the flow system's pressure, density, and entropy. 
The above equation can be written as 
\BA
\label{deriv}
\frac{\textrm{d}p^*}{\textrm{d}t} &=& \bigg(\frac{\P p^*}{\P\rho^*}\bigg)_S\frac{\textrm{d}\rho^*}{\textrm{d}t} 
+ \bigg(\frac{\P p^*}{\P S}\bigg)_{\rho^*}\frac{\textrm{d}S}{\textrm{d}t}\,.
\EA
The entropy is constant following the fluid particle as a thermodynamic state is in equilibrium, which implies 
$\frac{\textrm{d}S}{\textrm{d}t}=0$, thus Eq. (\ref{deriv}) reduced to
\BA
\frac{\textrm{d}p^*}{\textrm{d}t} &=& \bigg(\frac{\P p^*}{\P\rho^*}\bigg)_S\frac{\textrm{d}\rho^*}{\textrm{d}t}\,.
\EA
The fundamental thermodynamic state remains constant and is defined by the speed of sound, 
$a^*\equiv\gamma p^*/\rho^*$, and $\gamma$ represents the specific heats ratio.

The non-dimensional form of the governing equation will be obtained by using the following scaling 
\BA
\label{scale}
(x, y)\equiv (x^*, y^*)/L\,, \quad (u, v)\equiv (u^*, v^*)/U\,, \quad t\equiv t^*U/L\,, 
\quad \bar{u}\equiv\bar{u}^*/U\,, \quad p\equiv p^*/\rho^*U^2\,,
\EA
leads to
\BA
\label{non-dim-gov1}
\frac{\P u}{\P t} + \bar{u}\frac{\P u}{\P x} + v\frac{\P \bar{u}}{\P y} &=& -\frac{\P p}{\P x}\,,\\
\label{non-dim-gov2}
\frac{\P v}{\P t} + \bar{u}\frac{\P v}{\P x} &=& -\frac{\P p}{\P y}\,,\\
\label{non-dim-gov3}
M^2\bigg(\frac{\P p}{\P t} + \bar{u}\frac{\P p}{\P x}\bigg) + \frac{\P u}{\P x} + \frac{\P v}{\P y} &=& 0\,,
\EA
where Eqs. (\ref{non-dim-gov1}) and (\ref{non-dim-gov2}) are non-dimensional momentum equations. Combining 
the continuity equation and the entropy conservation yields Eq. (\ref{non-dim-gov3}). Here, $M\equiv U/a^*$ is 
the Mach number.

The boundary conditions are considered to be mathematically consistent and physically meaningful while also novel compared 
to classical studies of Blumen.\cite{Blumen70} Specifically, the base flow satisfies $U(y)=-1$ at $y=-h$ and $U(y)=1$ at 
$y=h$, over a bounded domain $y\in[-h, h]$. The normal velocity $v$ vanishes at the boundaries $y=\pm h$, enforcing 
impermeability. For pressure, a streamwise Newmann condition is enforced at the inlet and outlet to eliminate background 
pressure gradients, $\frac{\P p}{\P x}\Big|_{x=0, L}=0$. This framework facilitates capturing nonlinear saturation, 
bifurcation behavior, and the global dynamics of the perturbation amplitude through a Landau system, which is analyzed 
further using phase portraits and three-dimensional trajectories. The approach retains analytical tractability while 
incorporating compressibility and nonlinearity in a novel and generalizable way to a wider class of shear-driven 
instabilities.

\section{Weakly Nonlinear Stability Analysis\label{sec:wnsa}}
The weakly nonlinear stability analysis aims to capture the nonlinear effect that arises beyond the linearized equations 
and determine their impact on the evolution of disturbances. Here, we perform the weakly nonlinear stability analysis by 
perturbing the system of governing equations.

\subsection{Expand Variables in Perturbation Series\label{sec:sub1}}
Assume the flow variables (e.g., velocity, pressure, and density) are expanded in terms of a small perturbation parameter, 
representing the amplitude of the disturbance:
\BA
\label{expand}
\label{wnsa-per1}
{\bf u} &=& {\bf u}_0 + \varepsilon{\bf u}_1 + \varepsilon^2{\bf u}_2 + O(\varepsilon^3)\,,\\
\label{wnsa-per2}
\rho &=& \rho_0 + \varepsilon\rho_1 + \varepsilon^2\rho_2 + O(\varepsilon^3)\,,\\
\label{wnsa-per3}
p &=& p_0 + \varepsilon p_1 + \varepsilon^2 p_2 + O(\varepsilon^3)\,,
\EA
where $\varepsilon<<1$ is a small parameter representing the amplitude of the perturbations and (${\bf u}_0, \rho_0, p_0$) 
represents the base state (steady and independent of time\cite{Blumen70}), and higher-order terms (${\bf u}_1, \rho_1, p_1$ 
and (${\bf u}_2, \rho_2, p_2$) represent the first-order and second-order perturbations, respectively. 

Substituting (\ref{expand}) into Eqs. (\ref{non-dim-gov1})-(\ref{non-dim-gov3}), we collect equations in a different 
order of perturbations and solve the set of equations. At zeroth order $O(1)$, the solution yields $u_0=\bar{u}$, 
$v_0=0$, represents the steady, uniform base flow with velocity $\bar{u}$ in the $x-$direction. 

Similarly, the set of first-order perturbation equations becomes:
\BA
\label{non-dim-per1}
\frac{\P u_1}{\P t} + \bar{u}\frac{\P u_1}{\P x} &=& -\frac{\P p_1}{\P x}\,,\\
\label{non-dim-per2}
\frac{\P v_1}{\P t} + \bar{u}\frac{\P v_1}{\P x} &=& -\frac{\P p_1}{\P y}\,,\\
\label{non-dim-per3}
M^2\bigg(\frac{\P p_1}{\P t} + \bar{u}\frac{\P p_1}{\P x}\bigg) + \frac{\P u_1}{\P x} + \frac{\P v_1}{\P y} &=& 0\,,
\EA

We decompose the wave perturbation into normal modes, which is represented as
\BA
\label{normal-mode}
u_1 = \hat{u}_1e^{\i(k_xx + k_yy - \omega t)}\,, \qquad 
v_1 = \hat{v}_1e^{\i(k_xx + k_yy - \omega t)}\,, \qquad 
p_1 = \hat{p}_1e^{\i(k_xx + k_yy - \omega t)}\,,
\EA
where $k=\sqrt{k_x^2+k_y^2}$, $k_x$ and $k_y$ are the wavenumbers of the disturbance in the $x$- and $y$-directions and 
$\omega$ is the complex angular frequency with the complex wave velocity $c=\omega/k$. The growth rate $\sigma$ is related 
to the imaginary part of the angular frequency $\omega$ with a relation $\sigma=\Im(\omega)$. $\sigma$ represents the growth 
rate of small perturbations in a linear stability theory context. Thus, if $\sigma>0$, the perturbation grows exponentially, 
signifying instability, and if $\sigma<0$, the perturbation decays, signifying stability. $\sigma$ measures the growth (or 
decay) of small perturbations over time and determine the neutral stability curves.

Now, incorporating the normal mode decomposition (\ref{normal-mode}) into Eqs. (\ref{non-dim-per1})-(\ref{non-dim-per3}), 
for the first-order solutions, which yields the following algebraic system of equations:
\BA
\label{1st-order-eq1}
(\omega - k_x\bar{u})\hat{u}_1 &=& k_x\hat{p}_1\,,\\
\label{1st-order-eq2}
(\omega - k_x\bar{u})\hat{v}_1 &=& k_y\hat{p}_1\,,\\
\label{1st-order-eq3}
M^2(\omega - k_x\bar{u})\hat{p}_1 + k_x\hat{u}_1 + k_y\hat{v}_1 &=& 0\,.
\EA
Solving (\ref{1st-order-eq1}) and (\ref{1st-order-eq2}), we obtained the solutions for $\hat{u}_1$ and $\hat{v}_1$ as
\BA
\label{Sol-1-2}
\hat{u}_1 = \frac{k_x\hat{p}_1}{\omega - k_x\bar{u}}\,, \qquad \hat{v}_1 = \frac{k_y\hat{p}_1}{\omega - k_x\bar{u}}\,.
\EA
Substituting (\ref{Sol-1-2}) into (\ref{1st-order-eq3}), we get
\BA
\label{Sol-3}
M^2(\omega - k_x\bar{u})\hat{p}_1 + k_x\bigg(\frac{k_x\hat{p}_1}{\omega - k_x\bar{u}}\bigg) 
+ k_y\bigg(\frac{k_y\hat{p}_1}{\omega - k_x\bar{u}}\bigg) &=& 0\,.
\EA
From (\ref{Sol-3}), we attain the dispersion relation:
\BA
\label{Dis-1}
M^2(\omega - k_x\bar{u}) &=& k_x^2 + k_y^2\,.
\EA
Therefore, we attained two solutions of $\omega$ after solving (\ref{Dis-1}) as
\BA
\label{omega1}
\omega &=& k_x\bar{u} \pm\frac{\sqrt{k_x^2 + k_y^2}}{M}\,,
\EA
represents the frequencies of the upstream and downstream waves, respectively. The dispersion relation indicates that 
the system supports waves with two branches (depending on the $\pm$ sign). The term $\frac{\sqrt{k_x^2 + k_y^2}}{M}$ 
represents the effect of the compressibility on the wave speed, with a higher Mach number $M$ leading to slower wave 
propagation. Similarly, we solved for the second-order system of equations. At the second-order ($O(\varepsilon^2)$), 
the nonlinear interactions between the first-order terms will generate second-order corrections to the flow. So, we 
collect all the $O(\varepsilon^2)$ terms from the expanded governing equations. These equations typically describe 
the growth or modulation of the first-order disturbances over time. We mentioned the calculations to get the solutions 
from the second-order system of equations in detail in Appendix \ref{appendix-2Sol}. 
%%%%%%%%%%%%%%%%%%%%%%%%%%%%%%%%%%%%%%%%%%%%%%%%%%%%%%
\begin{figure}
\BC
\subfigure[$M=0.1$]{\label{fig-1a}% 
\includegraphics[width=0.5\textwidth]{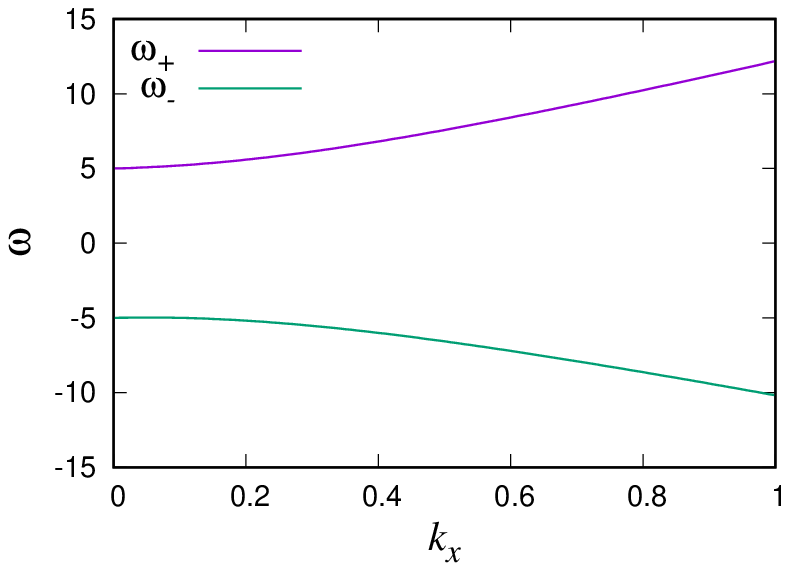}}\hfill%
\subfigure[$M=1.0$]{\label{fig-1b}% 
\includegraphics[width=0.5\textwidth]{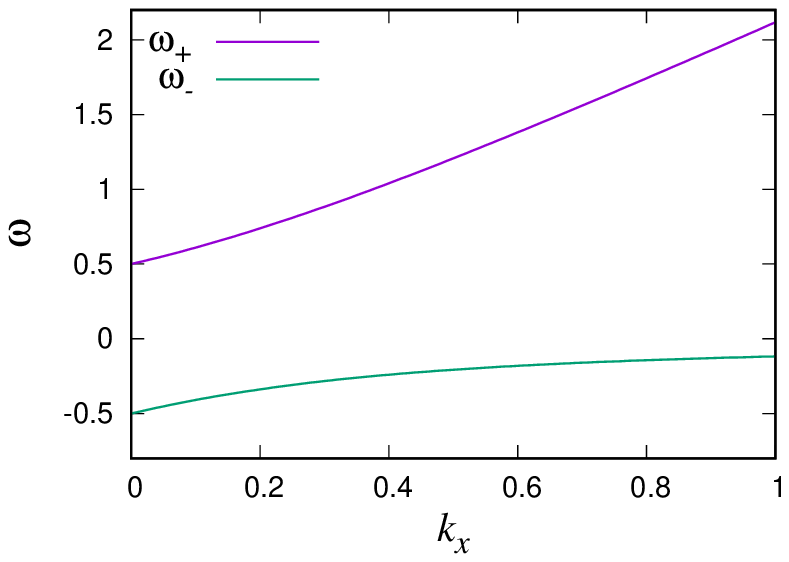}}
\caption{\label{fig:1} Wavenumber $k_x$ versus angular frequencies $\omega_+$ and $\omega_-$ is plotted with 
$M=0.1$ in (a) and $M=1.0$ in (b) illustrate the dispersion relation for the compressible system. Other values 
are fixed at $k_y=0.5$ and $\bar{u}=1.0$.}
\EC
\end{figure}
%%%%%%%%%%%%%%%%%%%%%%%%%%%%%%%%%%%%%%%%%%%%%%%%%%%%%%%%% 

In Figs. \ref{fig-1a} and \ref{fig-1b}, the two solutions for $\omega$ are shown for Mach number, $M=0.1$ and $M=1.0$, 
respectively. The figures illustrate the dispersion relation, depicting how the perturbation behaves in a compressible, 
moving fluid with varying Mach numbers. In this system, wave propagation occurs with a frequency $\omega$ determined 
by the wavenumbers $k_x$ and $k_y$. The term $k_x\bar{u}$ represents the Doppler shift due to the background flow in 
the $x$-direction. Specifically, when $k_x>0$, the wave frequency increases or decreases depending on whether the wave 
propagates with or against the flow. The frequency shift becomes more pronounced as the background velocity $\bar{u}$ 
increases. The separation of the two branches indicates that waves travel at varying speeds depending on their direction; 
either they propagate upstream ($\omega_-$) or downstream ($\omega_+$). This implies that waves moving downstream travel 
faster due to assistance from the background flow, whereas waves moving upstream experience a reduction in speed as 
they move against the flow. Since the frequency depends linearly on the wavenumber magnitude, the system exhibits 
non-dispersive behavior, explaining that all wavelengths (or wave components) travel at the same speed. This behavior 
is similar to sound waves, where compressibility effects dominate. Furthermore, when $M\ll 1$ (low Mach number, subsonic 
conditions), the term $\frac{1}{M}$ becomes large, and the system behaves like highly compressible acoustic waves, 
resulting in more symmetric branches (as shown in Fig. \ref{fig-1a}). In contrast, when $M\geq 1$ (high Mach number, 
near or supersonic conditions), the wave propagation is less influenced by compressibility (as shown in Fig. 
\ref{fig-1b}). Instead, the background advection term $k_x\bar{u}$ dominates, leading to a stronger Doppler shift and 
increased asymmetry between the two branches.  
%%%%%%%%%%%%%%%%%%%%%%%%%%%%%%%%%%%%%%%%%%%%%%%%%%%%%%
\begin{figure}
\BC
\subfigure[]{\label{fig-2a}% 
\includegraphics[width=0.5\textwidth]{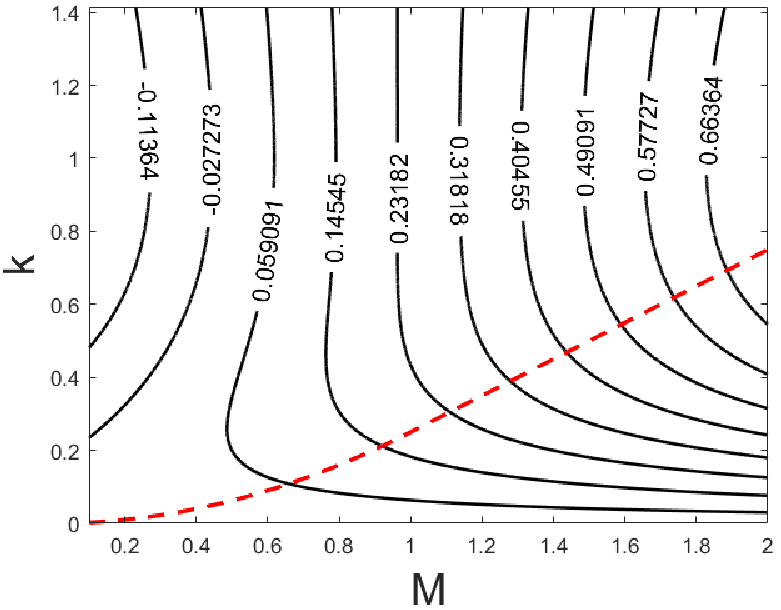}}\hfill%
\subfigure[]{\label{fig-2b}% 
\includegraphics[width=0.5\textwidth]{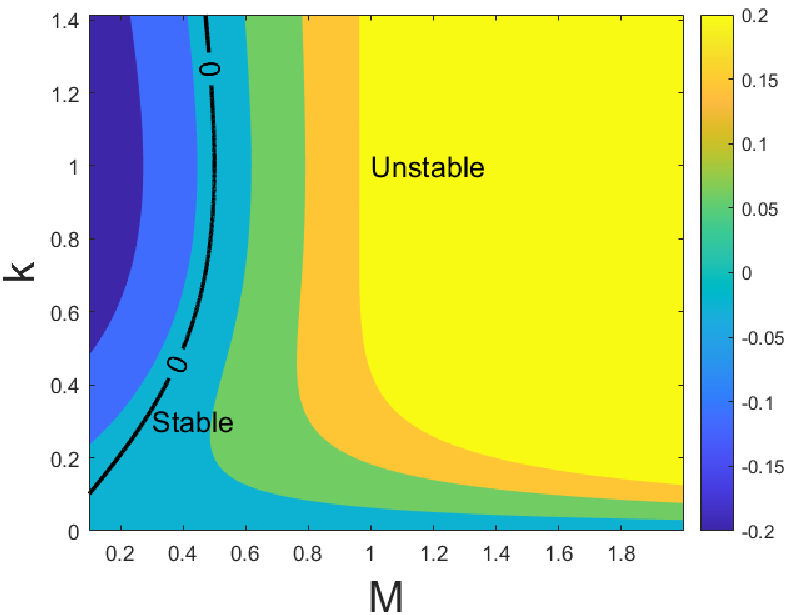}}
\caption{\label{fig:2} (a) Contour plot of the growth rate $\sigma$ in the ($M, k$) plane. The red dashed line 
indicates the locus of maximum growth rate for each Mach number $M$. (b) The colored contour map depicting regions of 
stability and instability based on the sign of $\sigma$, with the transition boundary separating stable and unstable 
zones marked.}
\EC
\end{figure}
%%%%%%%%%%%%%%%%%%%%%%%%%%%%%%%%%%%%%%%%%%%%%%%%%%%%%%%%%

Fig. \ref{fig-2a} illustrates the isoline of the growth rate $\sigma$ in the $M$, $k$ plane. The red dashed curve 
represents the maximum growth rate for each $M$, whereas $M$ increases, the wavenumber $k$ corresponding to the most 
unstable mode also increases. The isoline with the positive values of $\sigma$ is unstable, while the negative values 
imply stability. This is also shown in the contour Fig. \ref{fig-2b} with the neutral stability curve (solid line), 
where $\sigma=0$. The growth rate depends on both $k$ and $M$, indicating that the compressibility effects influence 
the instability behavior. As $M$ increases, the region of instability shifts towards higher wavenumbers, suggesting 
that the compressibility modifies the most unstable mode. This result directly impacts the transition to 
turbulence, particularly in high-speed shear layers, where the compressibility alters the dominant instability 
mechanisms. 
%%%%%%%%%%%%%%%%%%%%%%%%%%%%%%%%%%%%%%%%%%%%%%%%%%%%%%
\begin{figure}
\BC
\subfigure[Blumen\cite{Blumen70}]{\label{fig-3a}% 
\includegraphics[width=0.5\textwidth]{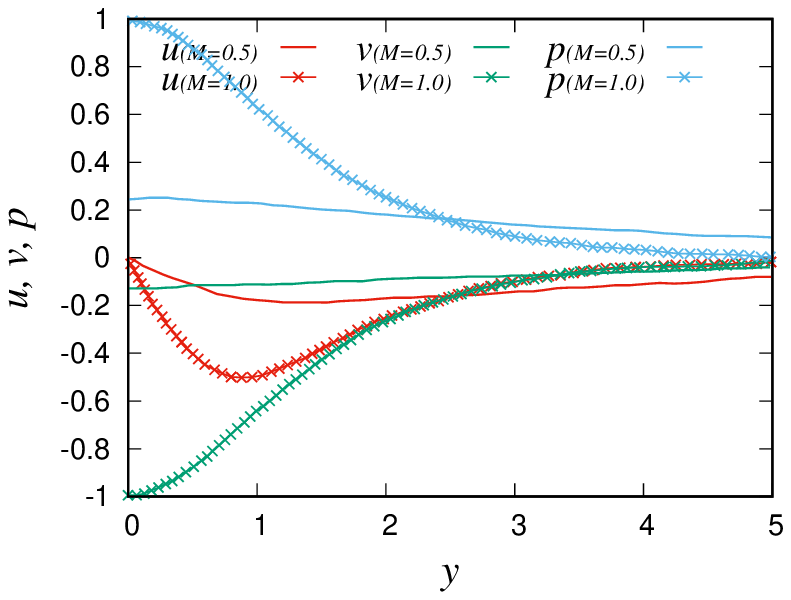}}\hfill%
\subfigure[Present study]{\label{fig-3b}% 
\includegraphics[width=0.5\textwidth]{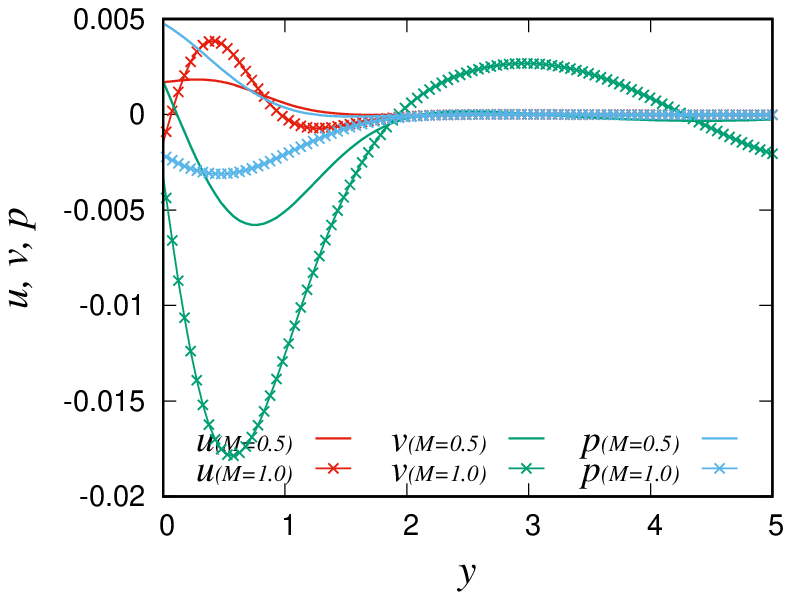}}
\caption{\label{fig:3} (a) Replotted from Blumen,\cite{Blumen70} showing neutral stability profiles of the 
velocity components $u$, $v$, and pressure $p$ as functions of $y$. (b) presents the corresponding neutral curves 
obtained from the present analysis. Solid and symbolized lines represent Mach numbers $M=0.5$ and $M=1.0$, respectively. 
Adapted from the results of W. Blumen, J. Fluid Mech. {\bf 40}, 769--781 (1970), for comparative purposes.}
\EC
\end{figure}
%%%%%%%%%%%%%%%%%%%%%%%%%%%%%%%%%%%%%%%%%%%%%%%%%%%%%%%%%

Figure \ref{fig:3} illustrates the neutral stability curves for the velocity components $u$, $v$, and pressure $p$ 
as a function of $y$ in a compressible shear flow. It includes comparing the results of Blumen's\cite{Blumen70} 
and the present study. In Fig. \ref{fig-3a}, Blumen's\cite{Blumen70} results show that the velocity components 
($u, v$) and pressure ($p$) perturbations exhibit relatively larger amplitudes, especially at a higher value of 
Mech number ($M=1.0$) is relatively high compared to the present study (Fig. \ref{fig-3b}), indicating that 
compressibility influences the stability characteristics by modifying shear instability. The lower magnitude of 
perturbations in our results suggests that instability mechanisms are altered when compressibility and different 
boundary conditions are considered, leading to a stabilizing effect due to the damping of perturbations. 
The maximum perturbation occurs near the center of the shear layer ($y\equiv 1$). The amplitude of $u(y)$ is lower 
in the present study (Fig. \ref{fig-3b}) compared to Blumen's (Fig. \ref{fig-3a}), suggesting that the boundary 
effect suppresses the growth of instability. For $M=1.0$, the perturbation magnitude is larger than for $M=0.5$, 
indicating that compressibility enhances streamwise velocity fluctuations, the transverse velocity fluctuations 
are even stronger at the center of the shear layer, and both become damp near the walls. The slight enhancement in 
the $v(y)$ magnitude at a high Mech number ($M=1.0$) is possibly due to the interaction between compressibility 
and shear-driven instabilities. Furthermore, the decay of pressure perturbations as $y$ increases aligns with 
the physical expectations of localized shear layer instability. The perturbation results in decay near the walls 
rather than growing indefinitely, indicating that Kelvin-Helmholtz instability behaves differently in a confined 
flow. This may affect aeroacoustics and flow control, where pressure feedback mechanisms in confined flows 
influence turbulence and instability growth. 
%%%%%%%%%%%%%%%%%%%%%%%%%%%%%%%%%%%%%%%%%%%%%%%%%%%%%%
\begin{figure}
\BC
\subfigure[Blumen\cite{Blumen70}]{\label{fig-4a}% 
\includegraphics[width=0.5\textwidth]{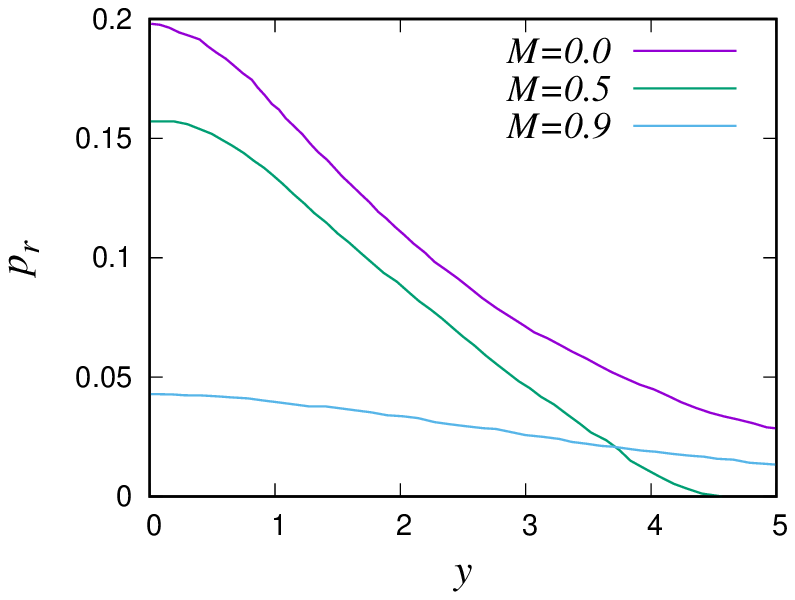}}\hfill%
\subfigure[Present study]{\label{fig-4b}% 
\includegraphics[width=0.5\textwidth]{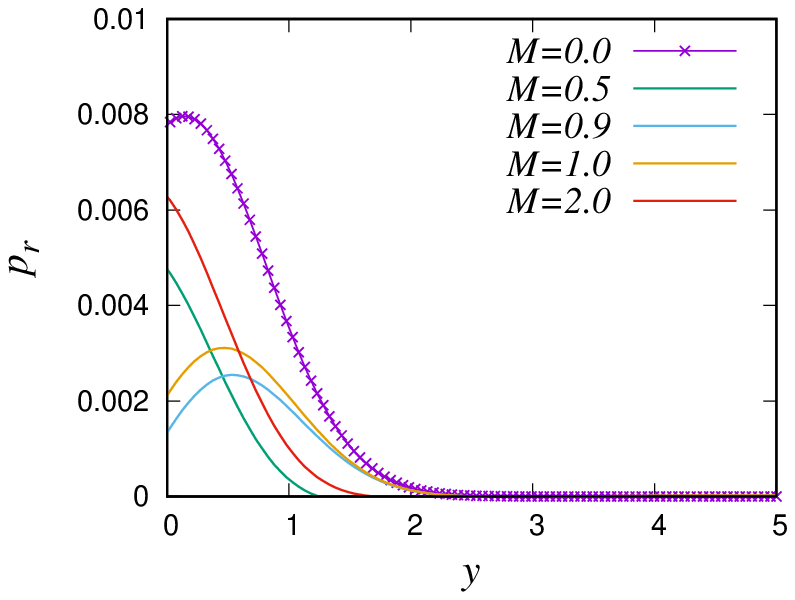}}\hfill%
\subfigure[Blumen\cite{Blumen70}]{\label{fig-4c}% 
\includegraphics[width=0.5\textwidth]{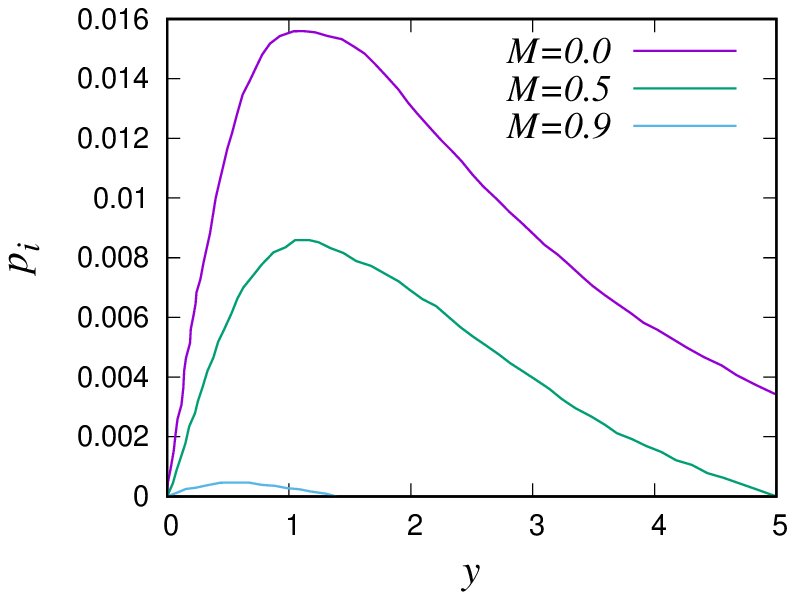}}\hfill%
\subfigure[Present study]{\label{fig-4d}% 
\includegraphics[width=0.5\textwidth]{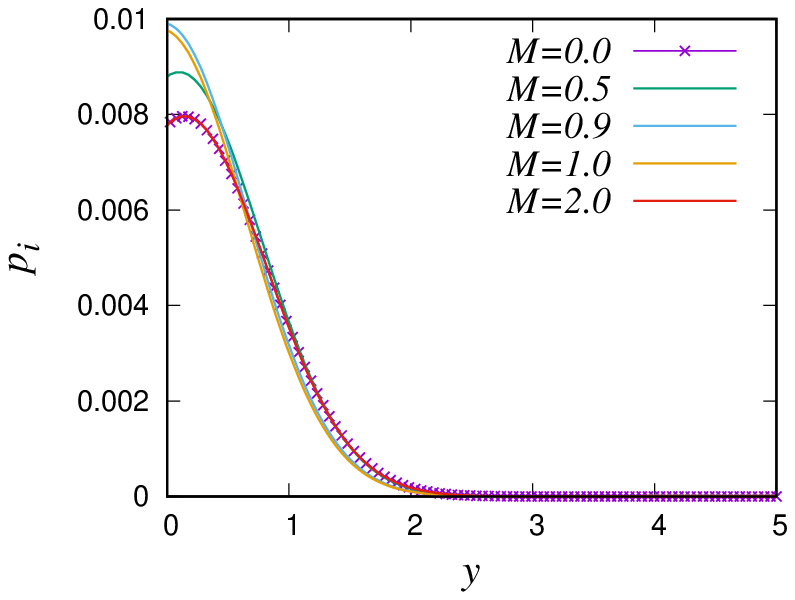}}
\caption{\label{fig:4} (a) and (c) show the real and imaginary parts of the pressure perturbation $p$ as 
functions of $y$ along the maximum growth rate curve, replotted from Blumen\cite{Blumen70} for selected Mach 
numbers. (b) and (d) present the corresponding results from the present study. Adapted from the results of 
W. Blumen, J. Fluid Mech. {\bf 40}, 769--781 (1970), for comparative purposes.}
\EC
\end{figure}
%%%%%%%%%%%%%%%%%%%%%%%%%%%%%%%%%%%%%%%%%%%%%%%%%%%%%%%%%

Figure \ref{fig:4} presents a comparison of the real ($p_r$) and imaginary ($p_i$) components of pressure 
perturbations as functions of $y$, with varying Mech numbers ($M$) in the present study and Blumen.\cite{Blumen70} 
Our study extended Blumens'\cite{Blumen70} work by including the higher Mach numbers ($M=1.0, 2.0$). In Figs. 
\ref{fig-4a} and \ref{fig-4c}, Blumens' results for $p_r$ and $p_i$ are shown, where different curves correspond to 
different Mach numbers ($M=0.0, 0.5, 0.9$). Whereas, Figs. \ref{fig-4b} and \ref{fig-4d} show our results, 
also incorporating a higher Mach number ($M=1.0, 2.0$), illustrating the behavior of $p_r$ and $p_i$ under stronger 
compressibility effects. The $p_r$ (a real component of pressure) represents the in-phase component of the pressure 
perturbation, which directly influences the velocity and vorticity dynamics in the shear layer. Meanwhile, the $p_i$ 
(an imaginary component of pressure) is associated with energy transfer mechanisms in the flow, influencing 
instability growth rates. Figure \ref{fig-4a} presents Blumen's results for $p_r$, where the amplitude of pressure 
perturbations decreases with increasing $M$, illustrating that the compressibility suppresses pressure fluctuations 
in the flow. These perturbations gradually decay with increasing $y$, showing that the pressure perturbations extend 
far from the shear layer. However, in the present study, the results initially follow the same decreasing trend with 
increasing $M$, indicating that compressibility stabilizes the instability, reducing the strength of pressure 
perturbations. Interestingly, the pressure perturbation amplifies for higher values of $M(\approx 1.0, 2.0)$, 
suggesting a non-monotonic compressibility effect and the possible occurrence of a secondary destabilization mechanism. 

In Fig. \ref{fig-4c}, Blumen's results show that as $M$ increases, the amplitude of $p_i$ (representing the growth 
rate of instability modes in the shear layer) decreases, indicating that compressibility stabilizes the instability. 
The peak location of $p_i(y)$ shifts slightly towards the centerline, indicating that the most unstable region moves 
inward as compressibility increases. However, the present study shows an opposite trend in Fig. \ref{fig-4d}; for 
increasing $M$ ($0\leq M\leq 0.9$), $p_i(y)$ increases, suggesting that moderate compressibility enhances 
instability growth. Furthermore, for higher values of $M(>0.9)$, $p_i(y)$ decreases, suggesting that perturbations 
become less effective at extracting energy from the base shear layer. This indicates that strong compressibility 
suppresses instability growth. Additionally, at high $M$, compressibility-driven instabilities reduce the energy 
transfer efficiency to the shear instability, decreasing the wave propagation speed.

Unlike Blumen's\cite{Blumen70} results, where perturbations spread out and weaken with increasing $M$, the present study 
reveals that pressure distributions become more intense and localized near the centerline. This indicates that 
compressibility amplifies rather than dampens instability, potentially due to stronger interactions between velocity 
and pressure fluctuations in a confined flow. At higher $M(>0.09)$, the amplitude trends in $p_r$ and $p_i$ differ. 
For $p_r$, compressibility injects more energy into the disturbance modes rather than stabilizing them, contradicting 
the classical notion that compressibility suppresses instability. This implies a different instability mechanisms 
may exist under the present boundary conditions. Moreover, for $p_i$, the system enters a regime where shear-driven 
instabilities weaken due to increased compressibility damping or modified flow structures. 

\subsection{Derive the Amplitude Equation\label{sec:sub2}}
To investigate the evolution of finite-amplitude perturbations near the onset of instability, we employ a weakly nonlinear 
stability analysis based on a multiple-scale expansion. Specifically, we introduce a slowly varying complex amplitude 
$A=A(T)$, where the slow time scale is defined as $T=\varepsilon^2 t$, and $\varepsilon\ll 1$ is a small parameter that 
measures the distance from the linear stability threshold. 

We begin by modifying the normal mode assumption to incorporate the amplitude modulation: 
\BA
\label{normal-mode-nsa}
[u_1, v_1, p_1] &=& [\hat{u}_1, \hat{v}_1, \hat{p}_1]A(T)e^{\i(k_xx + k_yy - \omega t)} + C.C.\,,
\EA
where $C.C.$ denotes the complex conjugate. The first-order solution represents the linear eigenfunction modulated by 
a slowly evolving amplitude.

Substituting (\ref{normal-mode-nsa}) into the governing equation with first-order perturbation terms will provide us 
the linearized problem at $O(\varepsilon)$, yielding the eigenvalue $\omega$ and eigenfunction 
($\hat{u}_1, \hat{v}_1, \hat{p}_1$).

Similarly, we modify the normal mode decomposition at $O(\varepsilon^2)$, having an quadratic interaction between first-order 
terms generate second harmonic and mean flow components, which yield: 
\BA
\label{normal-mode-2nd}
[u_2, v_2, p_2] &=& [\hat{u}_2, \hat{v}_2, \hat{p}_2]A(T)e^{2\i(k_xx + k_yy - \omega t)} + C.C. 
+ [\bar{u}_2, \bar{v}_2, \bar{p}_2]\,,
\EA
where $(\bar{u}_2, \bar{v}_2, \bar{p}_2)$ are the steady mean flow corrections.

Following the same trend, at $O(\varepsilon^3)$, the perturbation variables include the fundamental mode, second harmonics, 
and a mean flow correction: 
\BA
\label{normal-mode-3rd}
[u_3, v_3, p_3] &=& [\hat{u}_3, \hat{v}_3, \hat{p}_3]A^3(T)e^{3\i(k_xx + k_yy - \omega t)} 
+ [\tilde{u}_3, \tilde{v}_3, \tilde{p}_3]A(T)e^{\i(k_xx + k_yy - \omega t)} + [\bar{u}_3, \bar{v}_3, \bar{p}_3] + C.C.\,,
\EA
where $[\hat{u}_3, \hat{v}_3, \hat{p}_3]A^3(T)e^{3\i(k_xx + k_yy - \omega t)}$ represents the third harmonic, 
$[\tilde{u}_3, \tilde{v}_3, \tilde{p}_3]A(T)e^{\i(k_xx + k_yy - \omega t)}$ represents a resonant correction to the 
fundamental mode, and $(\bar{u}_3, \bar{v}_3, \bar{p}_3)$ represents the mean flow correction. The complex conjugate 
($C.C.$) terms ensure that the perturbation remains real. 

At $O(\varepsilon^3)$, the governing equations contain resonant terms that would cause unbounded growth unless a solvability 
condition is imposed. This condition is derived using Fredholm's alternative and involves the adjoint of the linear operator. 
The condition reads as: 
\BA
\label{WNS-sol1}
\int_{-h}^h \phi^*({N_1}_u + {N_1}_v + {N_1}_p) \textrm{d}y &=& 0\,,
\EA
where $\phi^*=(\phi_u^*, \phi_v^*, \phi_p^*)$, is the adjoint eigenfunction, and $({N_1}_u, {N_1}_v, {N_1}_p)$ represents 
the nonlinear terms with the known first- and second-order functions as coefficients, which in turn generate the resonant 
forcing components that must be canceled by the solvability condition, leading to an amplitude equation, which is commonly 
known as the Landau-Stuart equation, which governs the slow-time evolution of the perturbation amplitude $A(T)$:
\BA
\label{amplitude}
\frac{\textrm{d}A}{\textrm{d}T} &=& \mu A - \zeta |A|^2A\,,
\EA
where $\mu\in\mathbf{R}$ is a control parameter that refers to the linear growth/decay rate (obtained from linear stability 
analysis), and $\zeta\in\mathbf{C}$ refers to the nonlinear Landau coefficient, describing the nonlinear effects. The amplitude 
equation provides insight into the nonlinear stability of the flow, such as when $\mu>0$, the instability grows and leads 
to saturation due to nonlinear effects, and when $\mu<0$, the disturbances decay, and the flow remains stable. The sign of 
$\zeta$ determines whether the instability saturates to a steady state (supercritical bifurcation) or grows exponentially 
(subcritical bifurcation). Further discussion of our results is based on the estimated values of $\zeta$ and $\mu$ from 
Eq. (\ref{amplitude}), presented in Table \ref{tab:zetavalues}.

\section{Nonlinear dynamics and Bifurcation characteristics of compressible KHI\label{sec:bifurcation}}
%%%%%%%%%%%%%%%%%%%%%%%%%%%%%%%%%%%%%%%%%%%%%%%%%%%%%%
\begin{figure}
\BC
\subfigure[]{\label{fig-5a}% 
\includegraphics[width=0.5\textwidth]{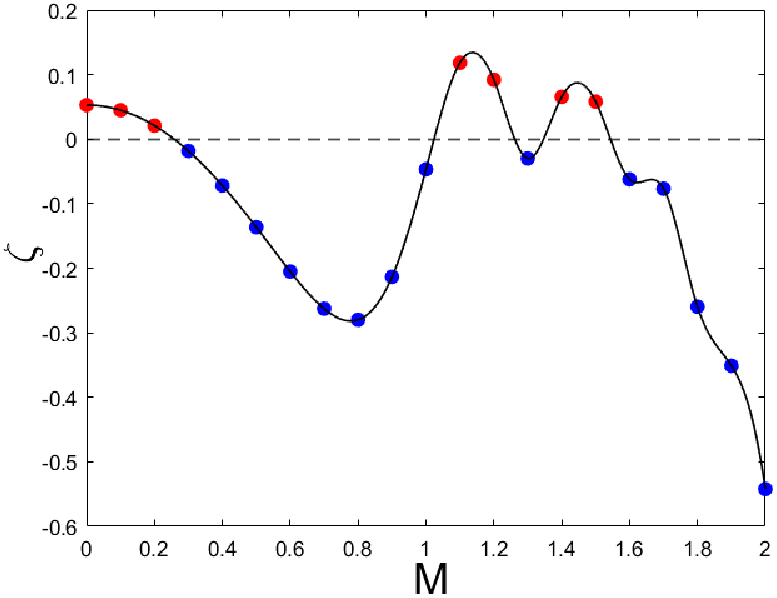}}\hfill%
\subfigure[]{\label{fig-5b}% 
\includegraphics[width=0.5\textwidth]{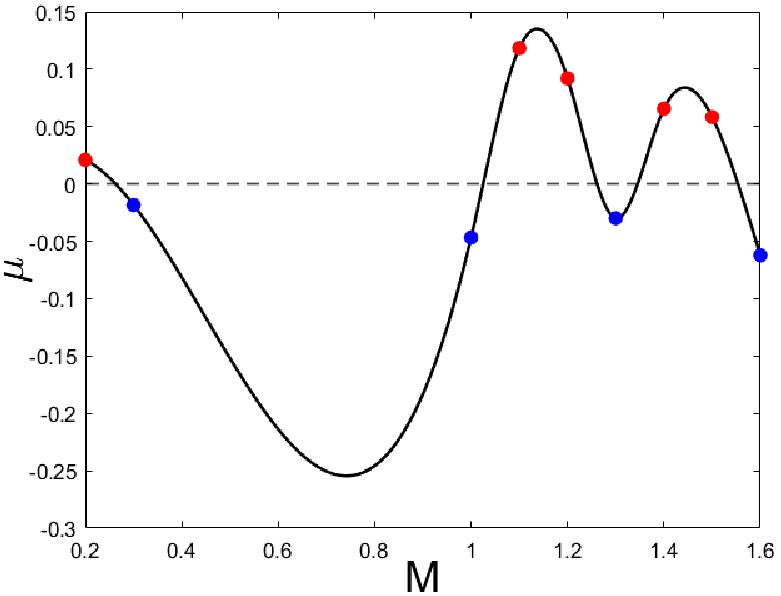}}\hfill%
\subfigure[]{\label{fig-5c}% 
\includegraphics[width=0.52\textwidth]{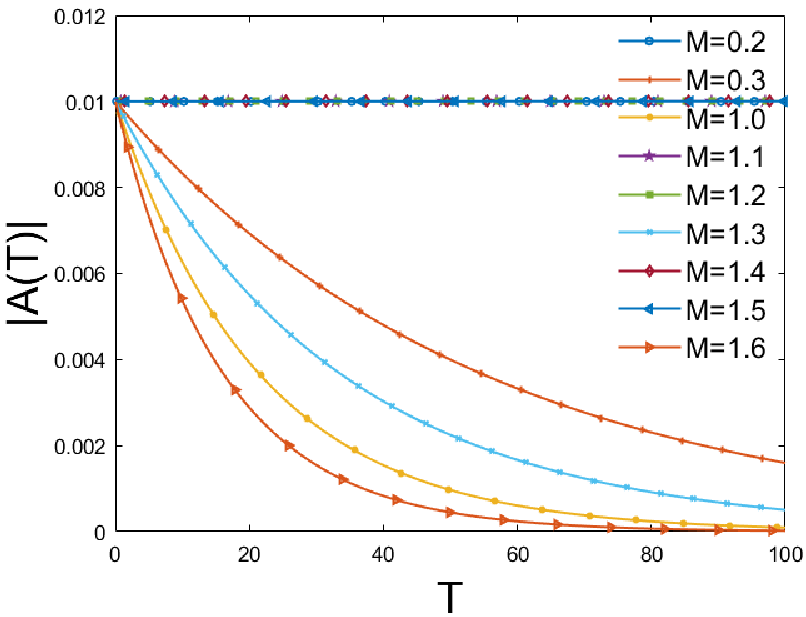}}\hfill%
\subfigure[]{\label{fig-5d}% 
\includegraphics[width=0.48\textwidth]{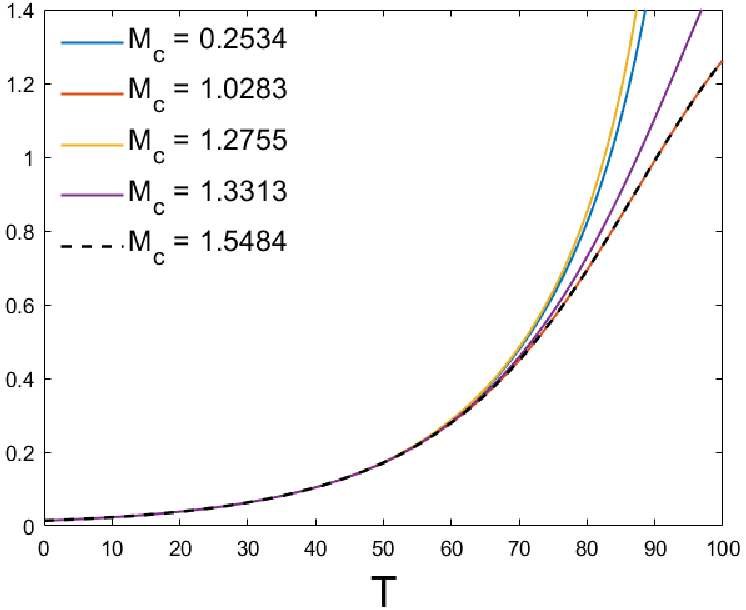}}
\caption{\label{fig:5} Bifurcation diagram illustrating the stable (blue dots) and unstable (red dots) points 
of $\zeta$ and $\mu$ in (a) and (b), respectively, with the variation of $M$. Saturation of amplitude $A(T)$ with time 
evolution, (c) with the variation of the critical Mach number, and (d) with the variation of the Mach number taken just 
before and after the critical Mach numbers.}
\EC
\end{figure}
%%%%%%%%%%%%%%%%%%%%%%%%%%%%%%%%%%%%%%%%%%%%%%%%%%%%%%%%%
To elucidate the interplay between compressibility and nonlinear instability mechanisms, we analyze the variation of 
the Landau coefficient $\zeta$, linear growth rate $\mu$, and the amplitude evolution $|A(T)|$ over a range of Mach 
numbers. These quantities offer a detailed picture of both the onset and the subsequent saturation of the 
Kelvin-Helmholtz instability (KHI) in compressible shear flows. 

In panels (a) and (b) of Fig. \ref{fig:5}, the results of the real part of the Landau coefficient 
$\Re(\zeta)$ and linear growth rate $\mu$ as a function of Mach number $M$ are illustrated, where the red 
and blue dots in Figs. \ref{fig-5a} and \ref{fig-5b}, refer to the stable and unstable $\zeta$ and $\mu$ points, 
respectively. The Positive $\Re(\zeta)$ value signifies nonlinear saturation, corresponding to supercritical 
bifurcations where the transitions smoothly from rest to finite-amplitude oscillations. In contrast, negative values 
indicate destabilizing nonlinearities, leading to subcritical bifurcations with the potential for finite-amplitude 
instabilities in linearly stable regimes. At low $M$ (weak compressibility), $\zeta>0$ dominates, indicating the flow 
is nonlinearly stable if unstable linearly. The system behaves like incompressible KHI, where the nonlinearity stabilizes 
the flow post-instability. As $M$ increases (intermediate $M$), $\Re(\zeta)$ oscillates in sign. There are 
non-monotonic transitions between $\zeta>0$ and $\zeta<0$, referring to the nonlinear behavior that switches between 
stabilizing and destabilizing. This is a profound showing that compressibility alters how energy is transferred 
between modes. Particularly, the local dips (around $M=1.05, 1.35$) suggest subcritical regimes ($\zeta<0$), where 
compressibility enhances nonlinear instability. Peaks near $M=1.25, 1.54$ indicate that nonlinear mechanisms recover 
their stabilizing nature. The results suggest that the compressibility introduces acoustic modes that can interact with 
shear modes nonlinearly, modifies the nature of the bifurcation, and the Mach number affects the strength and direction 
of energy transfer via nonlinear interactions due to the changes in multiple signs in $\Re(\zeta)$. The 
$\Re(\zeta)$ variation reflects a competition between shear-driven and compressibility-induced feedbacks. Near 
$M=1$, the phase speed of acoustic waves matches the base flow velocity, which causes resonance-like effects, leading 
to stronger energy feedback loops, possibly explaining why $\zeta$ becomes negative.

The linear growth $\mu>0$ and $\mu<0$ in Fig. \ref{fig-5b} indicates linearly unstable and stable. It shows multiple 
stability transitions with increasing $M$, such as when $0.25<M<1.0$ is stable, and beyond the range shows an unstable 
nature and continues re-stabilization and re-instability, suggesting nonlinear compressibility effects. The growth rate 
is non-monotonic due to the complex role of compressibility in altering the Kelvin-Helmholtz instability (KHI) mechanism. 
Some regimes with $\mu<0$ and $\zeta<0$ are nonlinearly unstable despite being linearly stable, showing classic subcritical 
behavior. These regimes define critical Mach numbers where the flow undergoes successive transitions, offering a physical 
basis for multiple bifurcation points observed in compressible KHI. Multiple critical values of the Mach number were found 
due to continuous changes in the sign. The critical values of $M$ are estimated by using linear interpolation between 
positive (or negative) and negative (or positive) two data points: 
\BA
\label{cricM}
M_c &=& M_{+(-)} + \Big(\frac{0-\Re(\zeta_{+(-)})}{\Re(\zeta_{-(+)})-\Re(\zeta_{+(-)})}(M_{-(+)} 
- M_{+(-)})\Big)\,.
\EA
Five critical values are computed as $M_c=0.2564, 1.0283, 1.2755, 1.3313, 1.5484$, which are plotted in Fig. \ref{fig-5d}. 
Figures \ref{fig-5c} and \ref{fig-5d} show the evolution of disturbance amplitude $|A(T)|$ over slow time $T$ for different 
Mach numbers and the critical Mach numbers, respectively. From Fig. \ref{fig-5c}, the amplitude curves for the stable Mach 
numbers (as shown in Fig. \ref{fig-5a}) are gradually decaying with the evolution of time. In contrast, for the unstable Mach 
numbers, all the amplitude curves are particularly beyond unity; the amplitude either remains flat or shoots up exponentially. 
In Fig. \ref{fig-5d}, the resultant curves focus on critical Mach number values around the transition between stability 
and instability. Here, all the curves exhibit nonlinear amplitude growth; however, with subtle variations in growth rate and 
saturation behavior. These results provide a richer view of the transition zone, suggesting a continuous bifurcation-like 
behavior rather than an abrupt shift. Unlike previous linear studies that analyzed only spectral stability, the present 
study employs a nonlinear Landau model to capture finite-amplitude effects and transient evolution. This approach uncovers 
complex dynamics near the stability boundary, including nonlinear saturation and mode competition. Identifying nuanced 
amplitude behaviors near the critical regime expands upon the previous foundational work,\cite{Blumen70,Blumen75,Drazin77} 
offering new insight into the nonlinear evolution of KHI in compressible fluids and its sensitivity to Mach number and 
spectral structure. These findings are particularly relevant to high-speed aerospace and astrophysical flows, where 
compressibility and finite-amplitude perturbations dominate transition and turbulence. These responses reflect the underlying 
energy exchange between the base flow, perturbation, and compressibility-modified wave modes. Each amplitude follows 
the trajectory: 
\BA
\label{AmpTrajectory}
|A(T)|^2 &=&\frac{\mu}{\Re(\zeta)}\big(1 - e^{-2\mu T}\big)^{-1}\,, \quad \mbox{if} \quad \mu>0\,. 
\EA 
From these results, we can physically interpret that the nonlinear effects saturate the amplitude when $\zeta>0$, even 
if the flow is linearly unstable. Moreover, saturation levels and time scales depend sensitively on both $\mu$ and $\zeta$. 
In addition, if $\zeta<0$, saturation may not occur, and the flow could experience finite-time blow-up or hysteresis.
%%%%%%%%%%%%%%%%%%%%%%%%%%%%%%%%%%%%%%%%%%%%%%%%%%%%%%%%%%%
\begin{table}
\BC
\begin{tabular}{cccc}
\hline
S.No. & $M$ & $\zeta$ & $\mu$\\
\hline
1 & 0.2 & 0.020982+0.42152\i & -0.1068\\
2 & 0.3 & -0.018339+0.40913\i & 0.0932\\
3 & 1.0 & -0.046715-0.19856\i &	-0.0566\\
4 & 1.1 & 0.11844-0.10808\i & 0.1434\\
5 & 1.2 & 0.092084+0.058672\i & -0.1510\\
6 & 1.3 & -0.029861+0.029111\i & 0.0490\\
7 & 1.4 & 0.065549-0.028316\i & 0.1374\\
8 & 1.5 & 0.058278+0.13589\i & -0.0968\\
9 & 1.6 & -0.062212+0.10415\i & 0.1032\\
\hline
\end{tabular}
\caption{\label{tab:zetavalues} Estimated values of $\zeta$ and $\mu$ for varying $M$.}
\EC
\end{table}
%%%%%%%%%%%%%%%%%%%%%%%%%%%%%%%%%%%%%%%%%%%%%%%%%

%%%%%%%%%%%%%%%%%%%%%%%%%%%%%%%%%%%%%%%%%%%%%%%%%%%%%%
\begin{figure}
\BC
\subfigure[$\zeta=0.020982+0.42152\i$]{\label{fig-6a}% 
\includegraphics[width=0.3\textwidth]{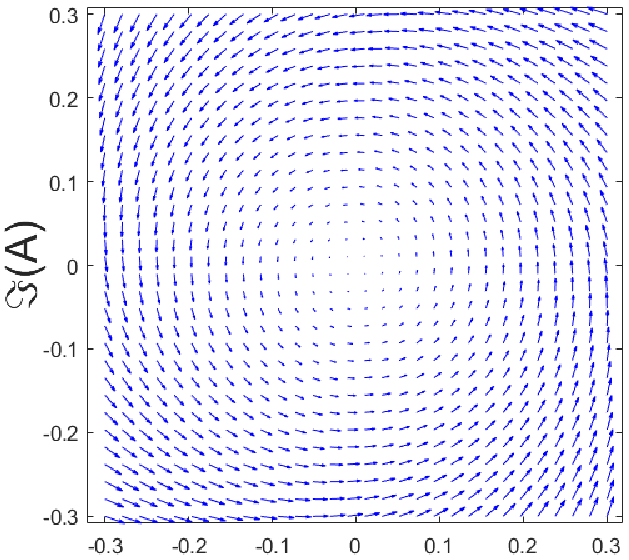}}\hfill%
\subfigure[$\zeta=-0.018339+0.40913\i$]{\label{fig-6b}% 
\includegraphics[width=0.275\textwidth]{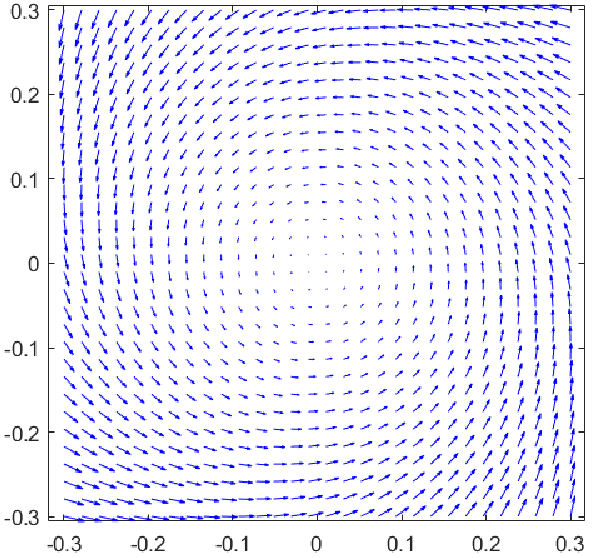}}\hfill%
\subfigure[$\zeta=-0.046715-0.19856\i$]{\label{fig-6c}% 
\includegraphics[width=0.275\textwidth]{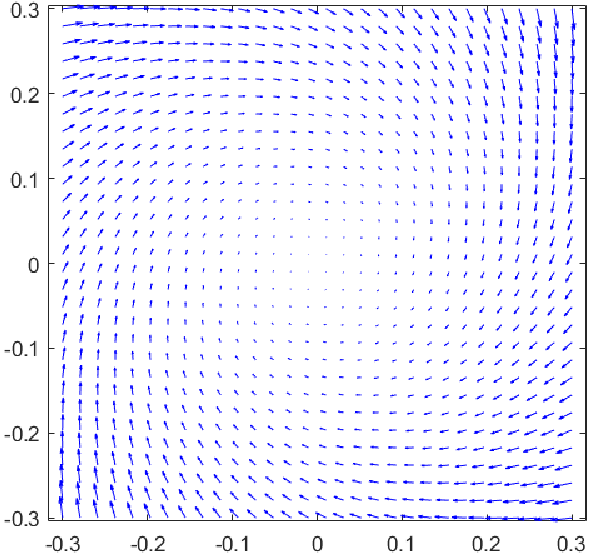}}\hfill%
\subfigure[$\zeta=0.11844-0.10808\i$]{\label{fig-6d}% 
\includegraphics[width=0.3\textwidth]{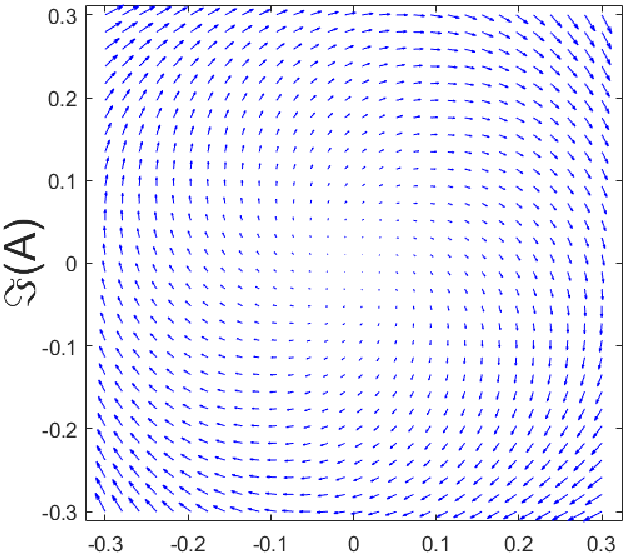}}\hfill%
\subfigure[$\zeta=0.092084+0.058672\i$]{\label{fig-6e}% 
\includegraphics[width=0.275\textwidth]{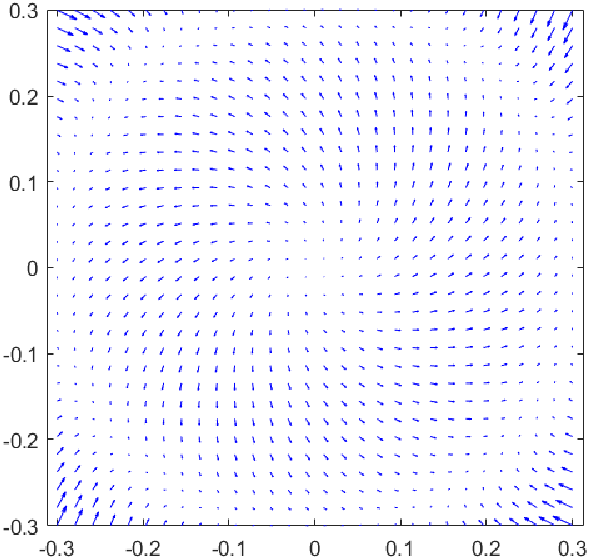}}\hfill%
\subfigure[$\zeta=-0.029861+0.029111\i$]{\label{fig-6f}% 
\includegraphics[width=0.275\textwidth]{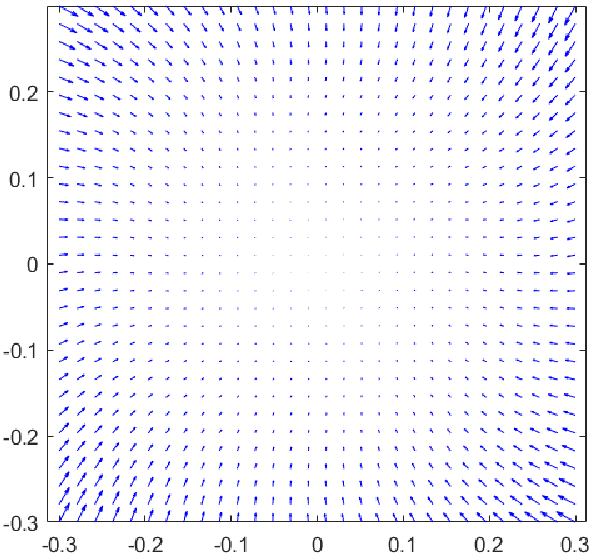}}\hfill%
\subfigure[$\zeta=0.065549-0.028316\i$]{\label{fig-6g}% 
\includegraphics[width=0.3\textwidth]{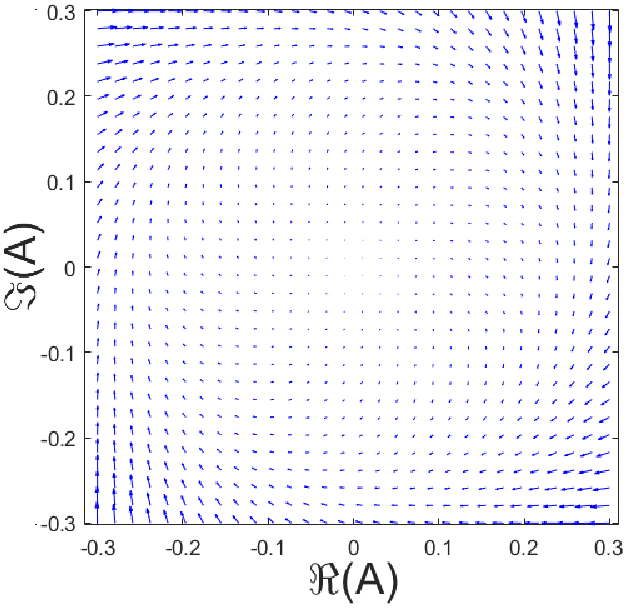}}\hfill%
\subfigure[$\zeta=0.058278+0.13589\i$]{\label{fig-6h}% 
\includegraphics[width=0.275\textwidth]{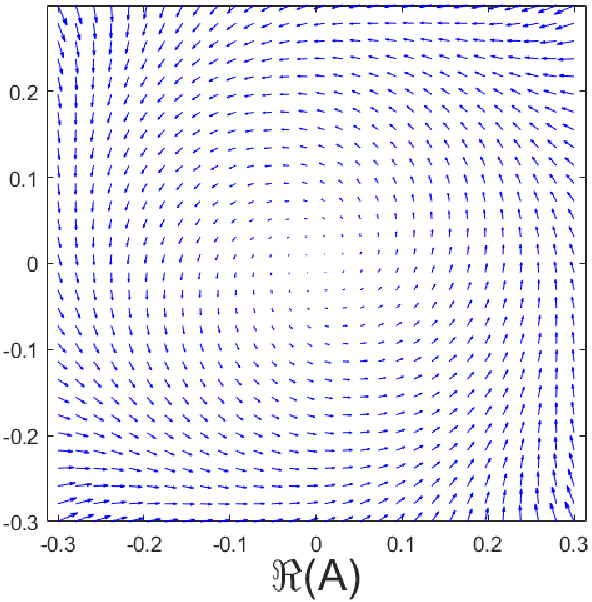}}\hfill%
\subfigure[$\zeta=-0.062212+0.10415\i$]{\label{fig-6i}% 
\includegraphics[width=0.275\textwidth]{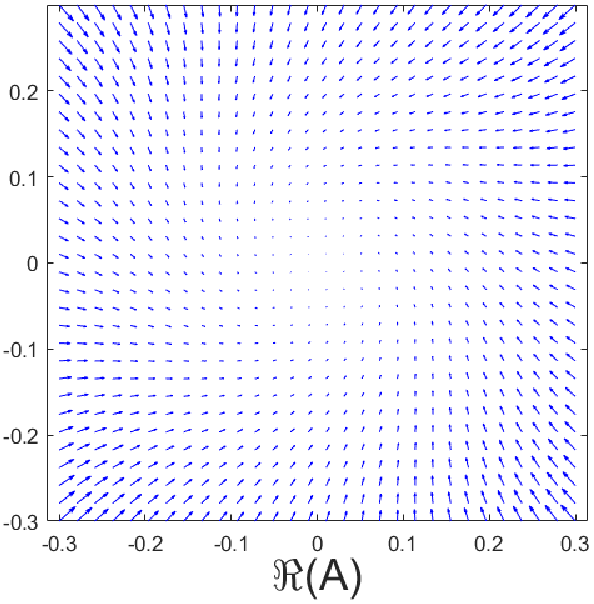}}
\caption{\label{fig:6}Phase portraits in the complex $A$-plane, with each subplot (a)-(i) illustrating the real 
and imaginary parts of the bifurcation amplitude for the critical values of Mach number $M$ (unstable) and its 
corresponding values of $\zeta$.}
\EC
\end{figure}
%%%%%%%%%%%%%%%%%%%%%%%%%%%%%%%%%%%%%%%%%%%%%%%%%%%%%%%%%
Figure \ref{fig:6} shows phase portraits in the complex $A$-plane, where $A$ represents the bifurcation amplitude 
arising from the weakly nonlinear analysis of a compressible shear flow. Each subfigure corresponds to the values 
just below and above the critical Mach number $M$ where the flow is nonlinearly unstable and shows how perturbations 
evolve in the complex amplitude plane, with $\Re(A)$ and $\Im(A)$ on the $x$- and $y$-axes respectively. The imaginary 
part of $\zeta$ introduces phase dynamics (rotating nature), while the real part determines growth versus saturation. 
In Fig. \ref{fig-6a}, since $\Re(\zeta)>0$, nonlinear effects saturate the instability. The positive imaginary part 
induces a rotational component. The system spirals show strong counter-clockwise rotation outward from the origin, 
indicating growing oscillations in both amplitude and phase. This represents a supercritical Hopf bifurcation in 
compressible flow where perturbations grow and saturate. In Fig. \ref{fig-6b}, the negative $\Re(\zeta)$ implies 
nonlinear destabilization with faster-growing perturbation due to nonlinear feedback. The system shows outward 
spiraling, even faster than panel (a), with strong phase rotation due to $\Im(\zeta)$. This may indicate subcritical 
instability where finite-amplitude disturbances destabilize the flow. In Fig. \ref{fig-6c}, both nonlinear growth 
and damping of rotational motion (negative $\Im(\zeta)$) are present. The phase rotation weakens while amplitude 
continues to grow unstably. This signifies a destabilized shear layer with decaying oscillation in phase but 
persistent growth in amplitude. In Fig. \ref{fig-6d}, positive $\Re(\zeta)$ allows saturation, while $\Im(\zeta)<0$ 
induces inward rotation. The flow evolves toward a limit cycle, possibly stabilizing at finite-amplitude oscillations, 
indicating saturation of linear instability and potential non-oscillatory steady-state. In Fig. \ref{fig-6e}, again 
$\Re(\zeta)<0$  with unstable focus and weakly spiraling outward means nonlinear terms reinforce instability. Flow 
perturbations amplify and rotate slowly. Physically, this suggests a compressible shear layer destabilizing even 
more strongly under perturbation, with a modest phase modulation. In Fig. \ref{fig-6f}, the nonlinear effect is weaker 
(smaller $|\zeta|$), so the dynamics are almost linear yet unstable. This demonstrates a delicate balance between 
growth and rotation. This might reflect the onset of chaotic or transitional behavior in weakly compressible regimes. In 
Fig. \ref{fig-6g}, the inward spiral implies decaying oscillations, and the system eventually settles into a non-zero 
amplitude steady-state. The system suggests nonlinearity stabilized the flow and compressibility-induced damping of 
instabilities. In Fig. \ref{fig-6h}, positive $\Re(\zeta)$ promotes saturation; positive $\Im(\zeta)$ gives sustained 
phase rotation, indicating a persistent oscillatory regime, where energy in perturbations remains bounded and rotates 
in the complex plane. It possibly describes nonlinear traveling waves in compressible shear. In Fig. \ref{fig-6i}, 
nonlinear term destabilizes ($\Re(\zeta)<0$), and $\Im(\zeta)>0$ gives phase rotation. The dynamics diverge from the 
origin, suggesting the growth of perturbation amplitude with oscillation, a hallmark of subcritical instability where 
even small disturbances can drive the transition. These figures reveal the route to transition and nonlinear wave 
evolution in compressible shear flow.

%%%%%%%%%%%%%%%%%%%%%%%%%%%%%%%%%%%%%%%%%%%%%%%%%%%%%%
\begin{figure}
\BC
\subfigure[$\zeta=0.020982+0.42152\i$]{\label{fig-7a}% 
\includegraphics[width=0.3\textwidth]{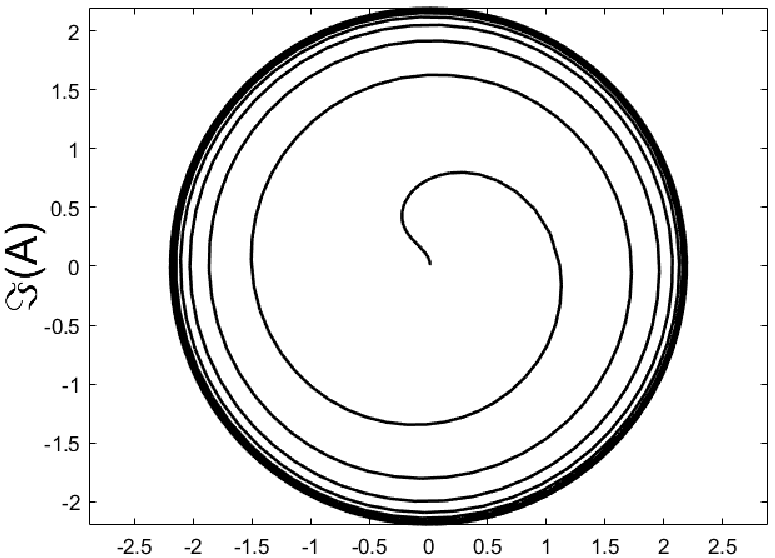}}\hfill%
\subfigure[$\zeta=-0.018339+0.40913\i$]{\label{fig-7b}% 
\includegraphics[width=0.3\textwidth]{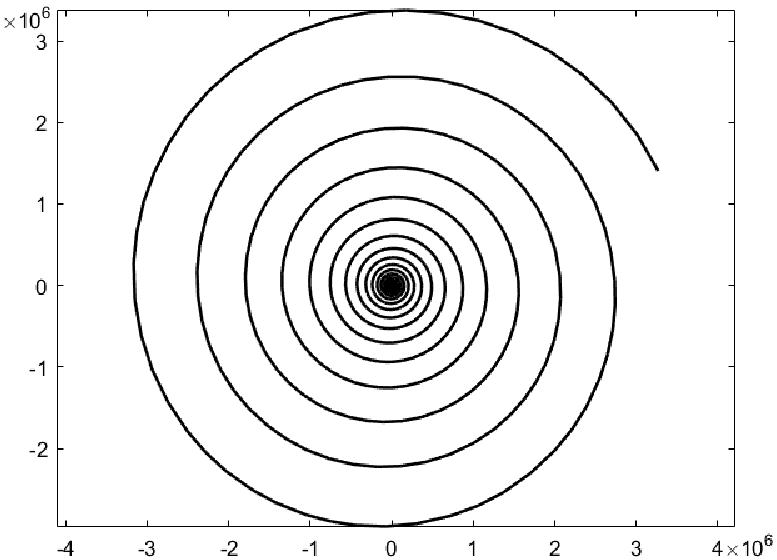}}\hfill%
\subfigure[$\zeta=-0.046715-0.19856\i$]{\label{fig-7c}% 
\includegraphics[width=0.29\textwidth]{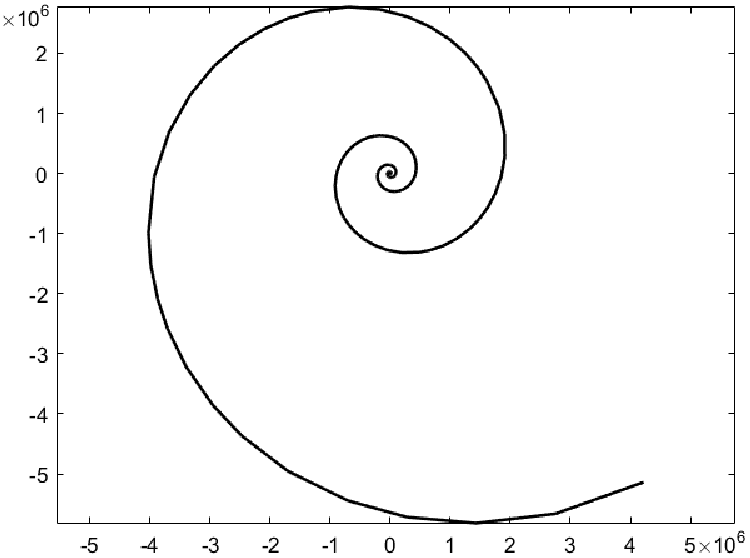}}\hfill%
\subfigure[$\zeta=0.11844-0.10808\i$]{\label{fig-7d}% 
\includegraphics[width=0.3\textwidth]{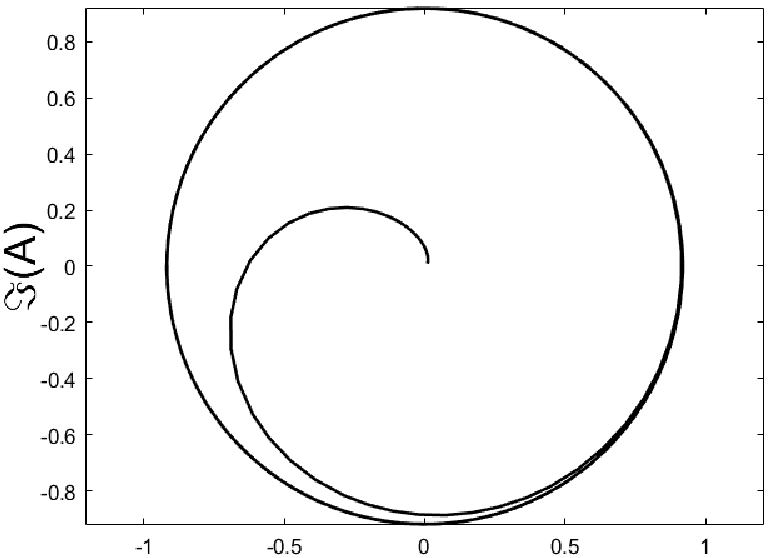}}\hfill%
\subfigure[$\zeta=0.092084+0.058672\i$]{\label{fig-7e}% 
\includegraphics[width=0.28\textwidth]{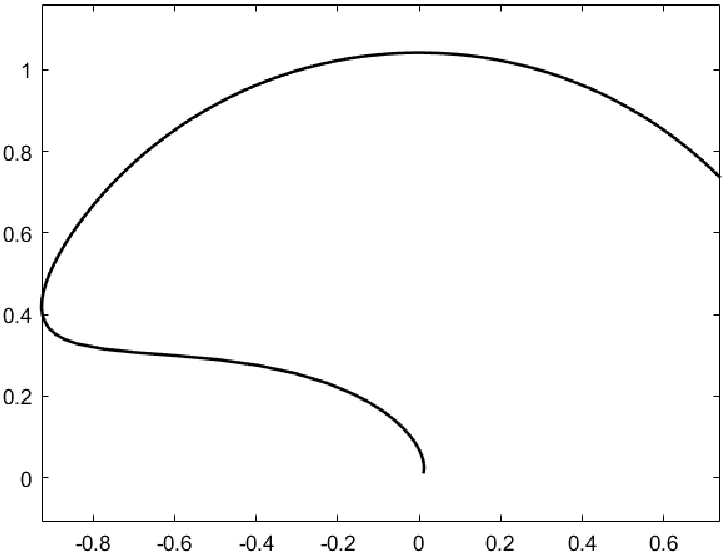}}\hfill%
\subfigure[$\zeta=-0.029861+0.029111\i$]{\label{fig-7f}% 
\includegraphics[width=0.29\textwidth]{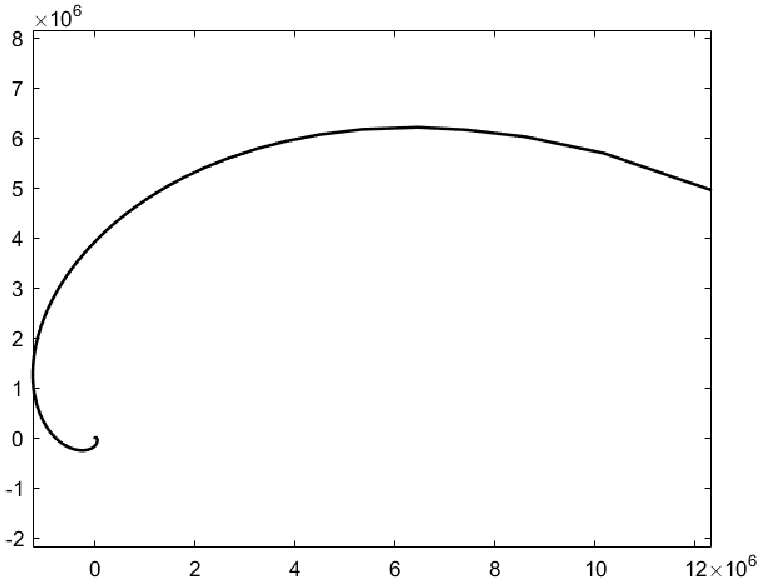}}\hfill%
\subfigure[$\zeta=0.065549-0.028316\i$]{\label{fig-7g}% 
\includegraphics[width=0.3\textwidth]{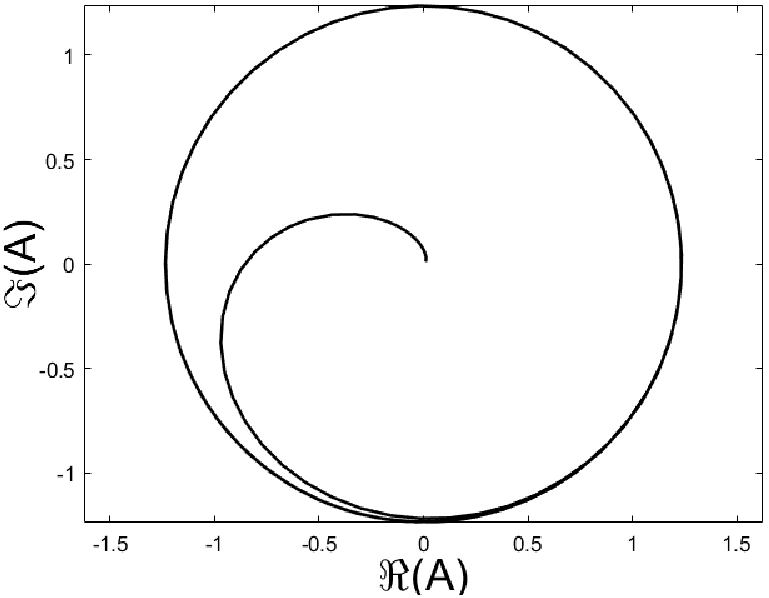}}\hfill%
\subfigure[$\zeta=0.058278+0.13589\i$]{\label{fig-7h}% 
\includegraphics[width=0.29\textwidth]{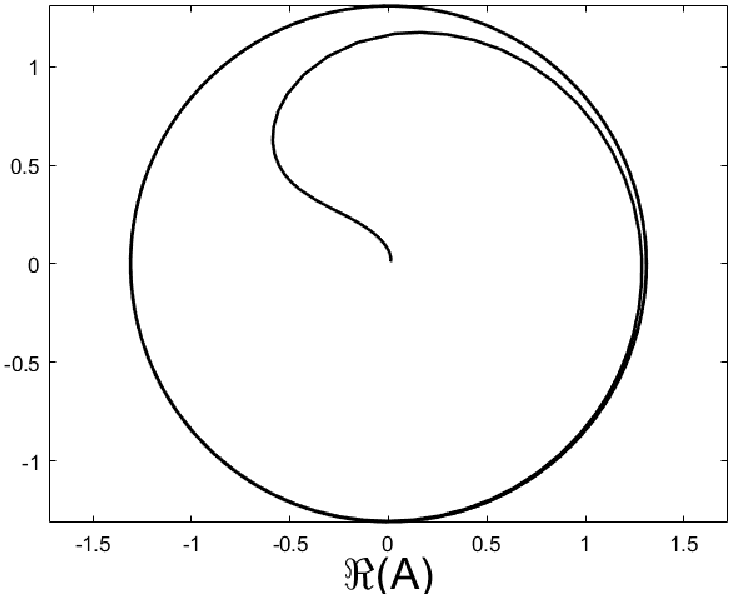}}\hfill%
\subfigure[$\zeta=-0.062212+0.10415\i$]{\label{fig-7i}% 
\includegraphics[width=0.29\textwidth]{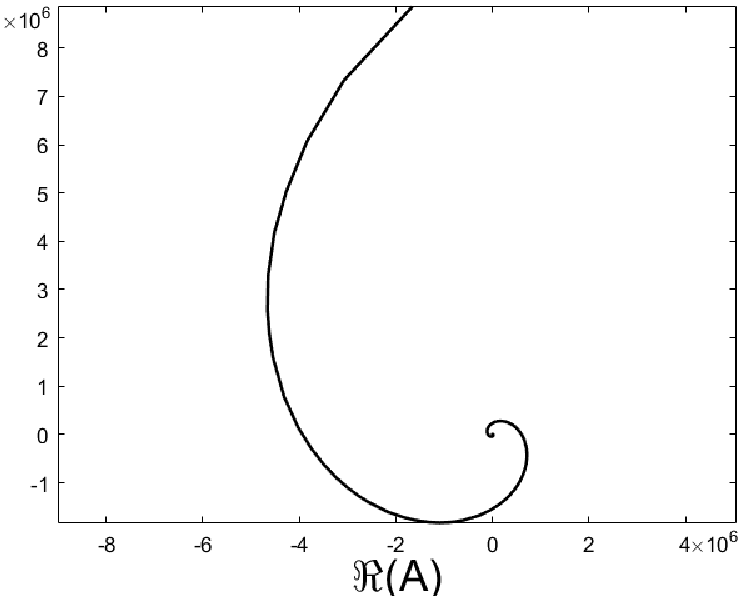}}
\caption{\label{fig:7}Two-dimensional phase portraits in the complex amplitude plane $A$ for a compressible 
Kelvin-Helmholtz instability system. Each subplot (a)-(i) corresponds to a distinct complex Landau coefficient 
$\zeta$, illustrating the nonlinear dynamics and stability characteristics of the amplitude evolution.}
\EC
\end{figure}
%%%%%%%%%%%%%%%%%%%%%%%%%%%%%%%%%%%%%%%%%%%%%%%%%%%%%%%%%
Figure \ref{fig:7} shows the two-dimensional phase portraits in the complex plane $A(=\Re(A)+\i\Im(A))$ for nine complex 
Landau coefficients $\zeta$ (corresponding to the below and above values of critical $M$) are significant for 
comprehending nonlinear dynamics and bifurcation behavior in the Kelvin-Helmholtz instability under compressible flow 
conditions. Each figure in Fig. \ref{fig:7} corresponds to a unique value of $\zeta$, showing the time evolution of 
perturbation amplitude $A(T)$ governed by the Landau-Stuart Eq. (\ref{amplitude}). The dynamics range from decaying 
spirals to limit cycles and nonlinear transitions, revealing the nature of bifurcation and nonlinear stabilization or 
destabilization mechanisms across different Mach numbers, $M$.

Figure \ref{fig-7a} illustrates that the system exhibits a dynamic response characterized by inward spirals that converge 
onto a finite-amplitude closed trajectory or limit cycle. This behavior reflects a transition from transient growth to 
a stable, periodic state and indicates nonlinearly saturated oscillations. Initially, perturbations grow due to 
linear instability mechanisms inherent in compressible shear flows; however, nonlinear effects (such as amplitude-dependent 
damping) eventually act to limit this growth, leading to a bounded, self-sustained oscillatory regime. This phenomenon 
is consistent with a supercritical Hopf bifurcation ($\Re(\zeta)>0$), where a stable limit cycle emerges smoothly 
as a control parameter (Mach number) exceeds a critical threshold. The balance between linear growth and nonlinear damping 
stabilizes the amplitude of oscillation. For stronger instabilities ($\Im(\zeta)$), the phase-space trajectory 
into the limit cycle exhibits more prominent spiraling due to nonlinear frequency modulation, wherein the oscillation 
frequency varies with amplitude. Such limit-cycle behavior is representative of a stable oscillatory shear layer, a flow 
feature commonly observed in compressible shear flows with moderate instability, such as those encountered in supersonic 
jets, shock-boundary layer interactions, and cavity flows. 

Figure \ref{fig-7b} illustrates that spiral trajectories characterize the system behavior in phase space that decays toward 
the origin, indicating the presence of a linearly stable fixed point. This suggests that any initial perturbations undergo 
exponential decay rather than amplifying or transitioning into sustained oscillations. The underlying dynamics reflect 
linear stability, wherein the real part of the dominant eigenvalue is sufficiently negative to suppress any transient 
growth. This stability persists across both negative values of the bifurcation parameter $\mu$ and small positive values, 
provided $\Re(\zeta)$ remains negative and sufficiently large in magnitude. Although the system may exhibit phase 
rotation due to complex eigenvalues, the overall energy of the perturbations decays over time, precluding the onset of 
sustained Kelvin-Helmholtz instabilities (KHI). From a physical standpoint, this behavior implies that the flow is resilient 
to perturbations, stability likely attributed to mechanisms such as dissipative compressibility effects known to dampen 
shear-driven instabilities in compressible flow regimes. Such dynamics are commonly observed in high-viscosity or strongly 
damped configurations, where the stabilizing influence of compressibility or diffusion dominates over inertial amplification 
mechanisms.

Figure \ref{fig-7c} explains that the system exhibits a dynamic behavior, referring to rapidly diverging outward spiral 
trajectories in phase space, indicative of an unstable regime characterized by nonlinear blow-up or explosive growth of 
perturbations. This behavior is consistent with a strong subcritical bifurcation, wherein $\Re(\zeta)$ remains 
negative, yet finite-amplitude perturbations experience runaway amplification due to the influence of nonlinear feedback 
mechanisms. In such cases, despite linear stability in the small amplitude limit, the system is highly susceptible to finite 
disturbances, which can grow uncontrollably and lead to a breakdown of coherent structures. This process often culminates in 
vortex pairing, flow reorganization, and a transition to turbulence. Physically, this instability is 
frequently observed in strongly sheared, compressible flows, particularly when the Mach number slightly exceeds a critical 
threshold. At this point, compressibility and shear interact to destabilize the flow nonlinearly, overcoming the dissipative 
effects that might otherwise maintain stability. The resulting explosive dynamics highlight the sensitivity of such systems 
to perturbations and underscore the importance of accounting for subcritical transition mechanisms in modeling and predicting 
turbulence onset in high-speed aerodynamic environments.

Figure \ref{fig-7d} explains that the system behavior is characterized by trajectories that converge onto a circular orbit 
in phase space, indicating a saturated limit cycle with a well-defined amplitude and frequency. This limit cycle reflects 
a self-regulated oscillatory state, wherein nonlinear energy transfer mechanisms balance any linear growth, leading to a 
dynamically stable configuration. A slight angular phase shift is observed in the orbital motion, attributed to a nonzero 
imaginary component of the growth rate, $\Im(\zeta)$, which governs the rotation in phase space. Specifically, a 
negative $\Im(\zeta)$ results in clockwise spiral evolution toward the limit cycle. This represents a canonical 
post-bifurcation state in shear-dominated compressible flows, emerging naturally after a supercritical Hopf bifurcation. 
Such dynamics are commonly associated with the formation of roll-up vortices that persist at constant amplitude, as seen 
in weakly unstable shear layers where compressibility and nonlinearity act to constrain the oscillatory behavior. A 
saturated, stable limit cycle implies that the flow, while dynamically active, remains bounded and predictable, 
an important feature in the study of flow control and transition prediction in compressible aerodynamic systems.

Figure \ref{fig-7e} explains that the system exhibits spiraling trajectories that saturate at a moderate amplitude, indicating 
a stable, slowly rotating limit cycle. This dynamic behavior suggests that perturbations initially grow before reaching a 
finite amplitude, at which point nonlinear effects act to stabilize the motion. The relatively weak nonlinearity in the system 
permits a gradual growth phase, allowing oscillations to develop before nonlinear saturation mechanisms limit further 
amplification. This behavior is characteristic of flows operating near the onset of instability, where the system is close 
to a bifurcation threshold but has not yet transitioned into strongly nonlinear or turbulent regimes. In such conditions, 
compressibility effects play a moderating role, effectively softening the system's response to perturbations by distributing 
energy across pressure and density fluctuations. This leads to a more subdued evolution than highly unstable configurations. 
It is commonly observed in mildly sheared compressible flows, such as those encountered in subsonic or transonic boundary 
layers and free shear layers near their critical conditions. The resulting dynamics are essential for understanding early-stage 
instability development and provide insights into the mechanisms that delay or control transition in compressible aerodynamic 
environments.

Figure \ref{fig-7f} explains a dynamical behavior characterized by outwardly growing spiral trajectories, signaling a weak 
yet persistent nonlinear blow-up. Although $\Re(\zeta)$ suggests linear stability, the system becomes nonlinearly 
unstable due to a subcritical bifurcation. In this regime, small perturbations would appear to decay under linear stability 
analysis, but finite-amplitude disturbances can trigger unbounded growth by overcoming the stabilizing effect predicted by 
linear theory. This form of dangerous instability is subtle and easily overlooked in traditional linear stability frameworks 
but can lead to dramatic consequences such as vortex merging or the early onset of turbulence. Such behavior is especially 
relevant in transitional compressible mixing layers, where small nonlinearities are amplified in the presence of 
compressibility and shear, facilitating a transition from orderly to chaotic flow states even at subcritical conditions. 
Recognizing and modeling these nonlinear dynamics is crucial for predicting and controlling transition phenomena in 
high-speed aerodynamic systems.

Figure \ref{fig-7g} exhibits a smooth convergence toward a circular orbit in the complex amplitude plane, reflecting the 
emergence of a stable limit cycle governed by low-frequency oscillations. The eigenvalue $\zeta=0.065549-0.028316\i$ reveals 
a small imaginary component, resulting in slow angular rotation and, thus, minimal frequency modulation. This behavior 
indicates a case of nonlinear saturation where the flow reaches a steady-state oscillatory regime following an initial growth 
phase. The smooth nature of the convergence suggests that nonlinear effects effectively counterbalance the weak linear 
instability, leading to a well-regulated amplitude and phase. This scenario is representative of a mildly unstable Kelvin-
Helmholtz instability (KHI) mode, which--while linearly unstable--is brought into a stable nonlinear equilibrium without 
transitioning to turbulence. Such dynamics are commonly encountered in compressible shear layers near their instability 
threshold, where compressibility prevents runaway growth. This case is a canonical example of how nonlinearity 
stabilizes low-intensity instabilities, yielding a predictable and sustained oscillatory structure in transitional 
compressible flows.

Figure \ref{fig-7h} exhibits a converging spiral trajectory in the complex amplitude plane, indicating a nonlinearly 
saturated limit cycle with prominent rotational dynamics. The system evolves from an initial transient toward a bounded, 
periodic state, where the amplitude stabilizes while the phase undergoes strong angular rotation. The governing eigenvalue, 
$\zeta$ has a relatively large imaginary component corresponding to a higher oscillation frequency. This elevated 
$\Im(\zeta)$ results in a faster rotational rate in phase space, reflecting a moderate frequency-modulated 
oscillatory mode. Such behavior suggests that, while the system remains stable due to nonlinear saturation, the dynamics 
are influenced by compressibility effects, particularly those associated with acoustic feedback mechanisms. In compressible 
shear flows, especially near transitional regimes, feedback between hydrodynamic instabilities and acoustic waves can 
modify both the frequency and stability of oscillations. The resulting limit cycle captures these interactions, revealing 
a flow regime where compressibility regulates amplitude and significantly alters temporal characteristics.

Figure \ref{fig-7i} exhibits a strikingly rapid outward spiral trajectory, indicating a strong nonlinear blow-up, despite 
the eigenvalue $\zeta$ suggesting linear stability due to a negative real part. This case exemplifies an energetic subcritical 
instability where nonlinear mechanisms amplify finite-amplitude perturbations uncontrollably, ultimately overwhelming the 
damping predicted by linear theory. This represents the most dangerous scenario among the cases presented, as the flow 
appears stable under small disturbances but becomes violently unstable under finite perturbations, an outcome invisible 
to traditional linear stability analysis. Physically, such behavior is characteristic of intense shear layer breakdown, 
often accompanied by vortex merging, roll-up instabilities, and the onset of turbulence, particularly in high Mech number 
compressible flows. In these regimes, the interplay between compressibility, nonlinear mode coupling, and energy transfer 
accelerates the divergence process, posing challenges for prediction and control. The dynamics captured here highlight 
the critical need to account for nonlinear effects when evaluating the stability of compressible shear flows, especially 
near transition thresholds where subcritical pathways to turbulence are dominant.

%%%%%%%%%%%%%%%%%%%%%%%%%%%%%%%%%%%%%%%%%%%%%%%%%%%%%%
\begin{figure}
\BC
\subfigure[$\zeta=0.020982+0.42152\i$]{\label{fig-8a}% 
\includegraphics[width=0.3\textwidth]{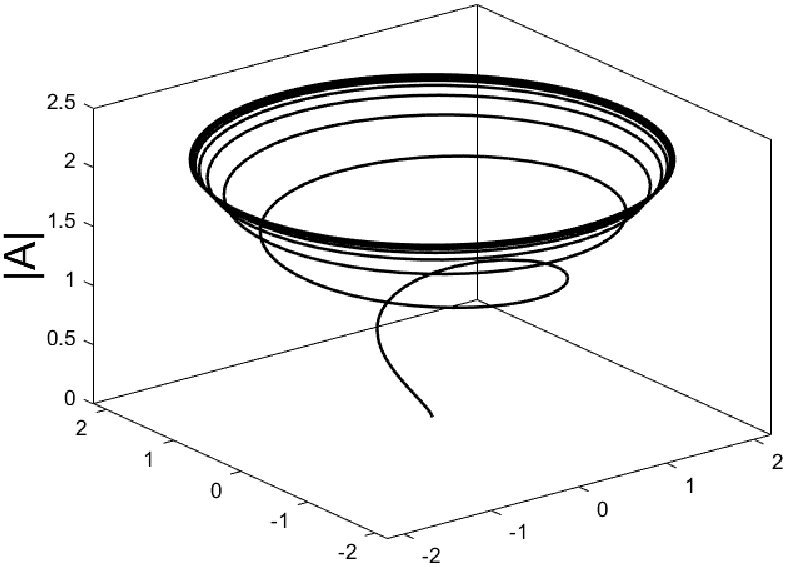}}\hfill%
\subfigure[$\zeta=-0.018339+0.40913\i$]{\label{fig-8b}% 
\includegraphics[width=0.28\textwidth]{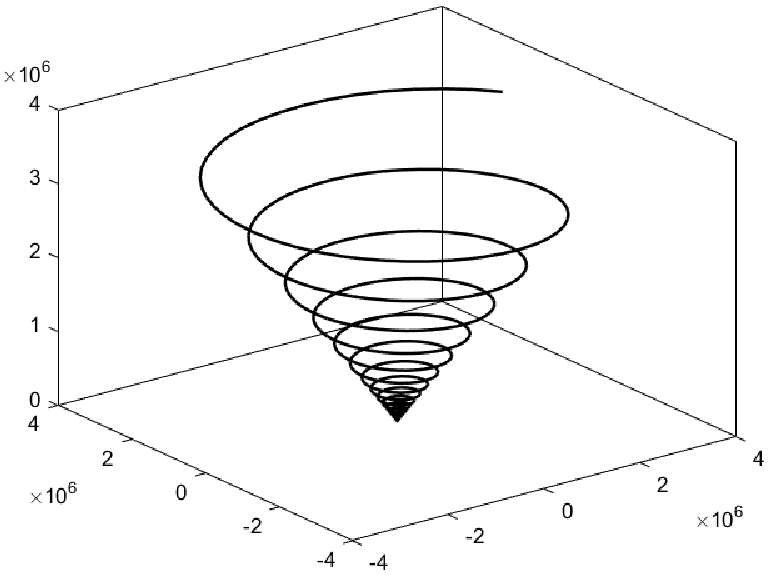}}\hfill%
\subfigure[$\zeta=-0.046715-0.19856\i$]{\label{fig-8c}% 
\includegraphics[width=0.28\textwidth]{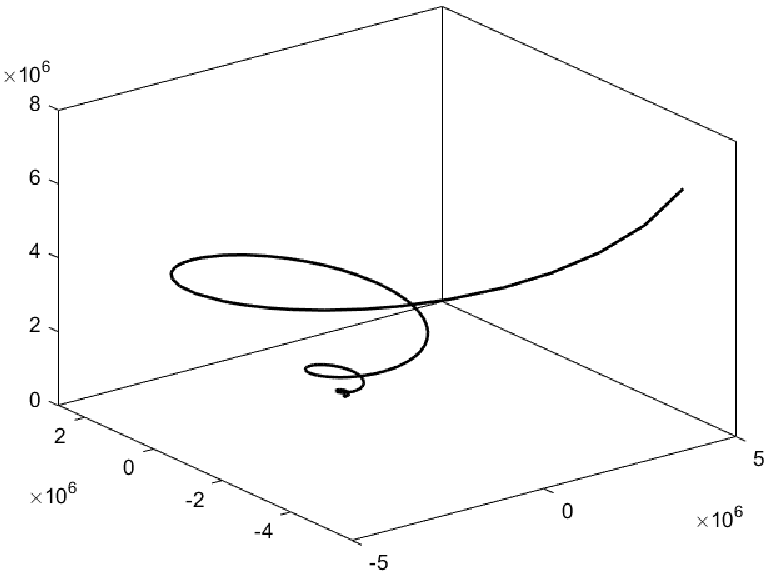}}\hfill%
\subfigure[$\zeta=0.11844-0.10808\i$]{\label{fig-8d}% 
\includegraphics[width=0.3\textwidth]{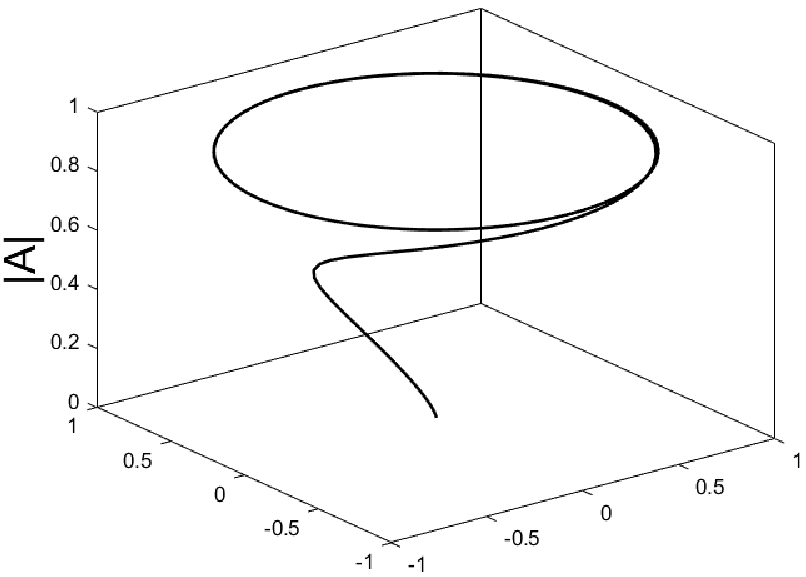}}\hfill%
\subfigure[$\zeta=0.092084+0.058672\i$]{\label{fig-8e}% 
\includegraphics[width=0.28\textwidth]{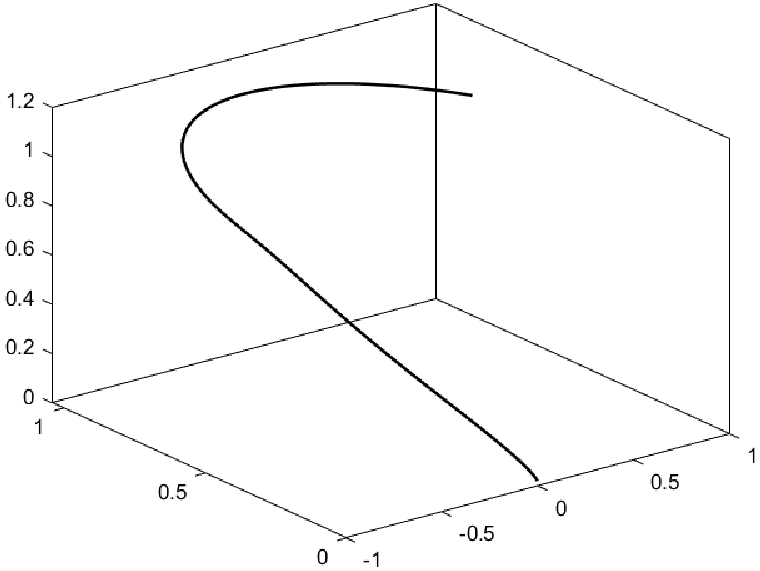}}\hfill%
\subfigure[$\zeta=-0.029861+0.029111\i$]{\label{fig-8f}% 
\includegraphics[width=0.28\textwidth]{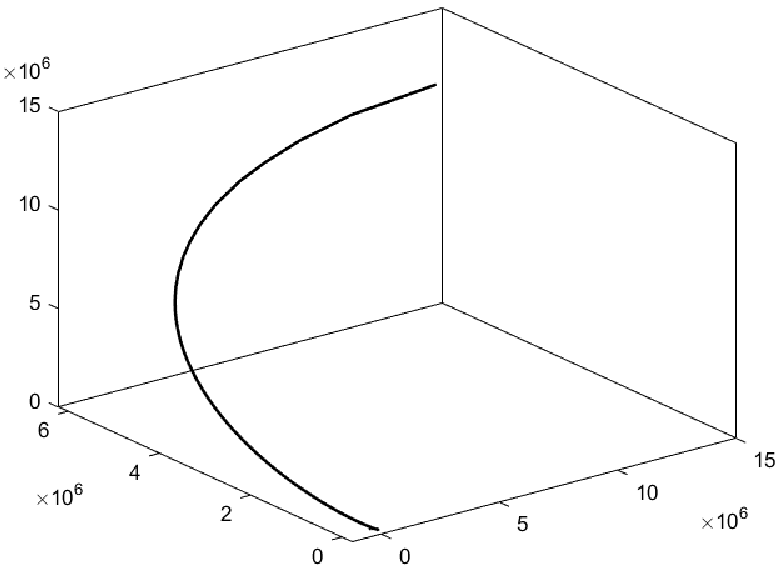}}\hfill%
\subfigure[$\zeta=0.065549-0.028316\i$]{\label{fig-8g}% 
\includegraphics[width=0.3\textwidth]{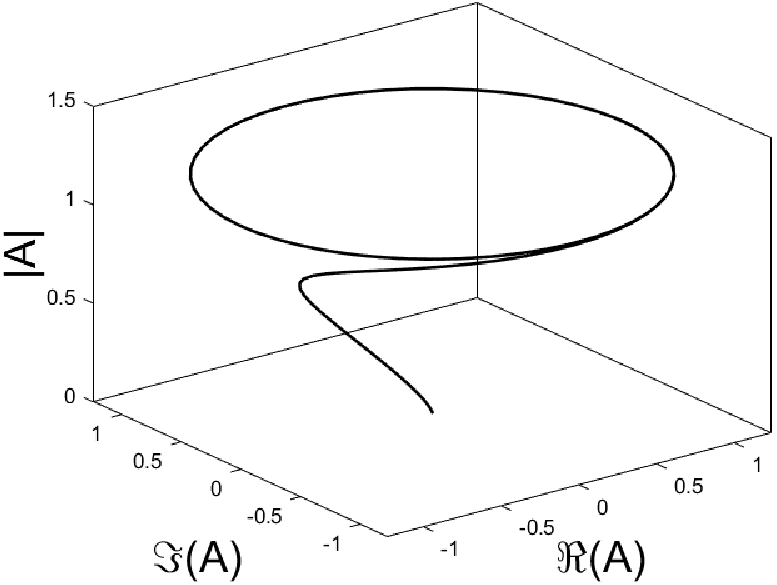}}\hfill%
\subfigure[$\zeta=0.058278+0.13589\i$]{\label{fig-8h}% 
\includegraphics[width=0.28\textwidth]{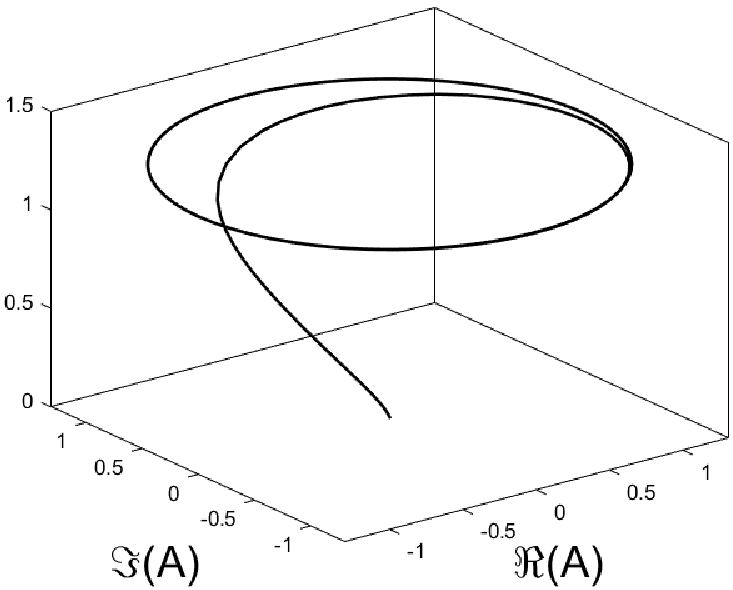}}\hfill%
\subfigure[$\zeta=-0.062212+0.10415\i$]{\label{fig-8i}% 
\includegraphics[width=0.28\textwidth]{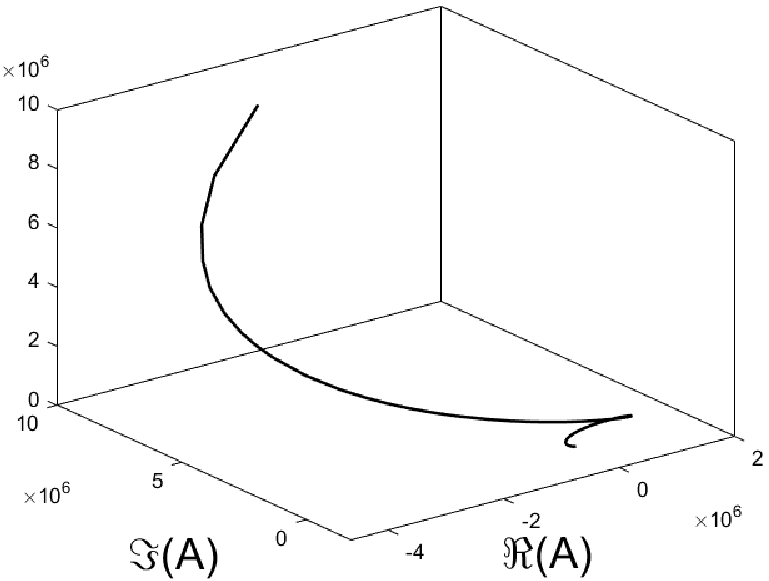}}
\caption{\label{fig:8} Three-dimensional phase-space trajectories of the complex amplitude $A(T)$ in the Landau 
system, plotted in the space of ($\Re(A), \Im(A), |A|$) for varying complex Landau coefficients 
$\zeta$, corresponding to different Mach numbers. Each subfigure (a)-(i) shows the nonlinear evolution of the 
perturbation amplitude $A(T)$, with the nature of the trajectories governed by the real and imaginary parts of 
$\zeta$.}
\EC
\end{figure}
%%%%%%%%%%%%%%%%%%%%%%%%%%%%%%%%%%%%%%%%%%%%%%%%%%%%%%%%%
Figure \ref{fig:8} represents three-dimensional trajectories of $A(T)\in\mathbf{C}$, with time evolving amplitude 
$|A|$, and behavior dictated by the Landau Eq. (\ref{amplitude}). The dynamics are controlled by the real and 
imaginary parts of $\zeta$ (nonlinear coefficient), with fixed $\mu$. In Fig. \ref{fig-8a}, where both $\Re(\zeta)>0$ 
and $\Im(\zeta)>0$ interprets nonlinear saturation and oscillatory behavior and exhibits supercritical Hopf bifurcation. 
The trajectory shows the spiral outward and saturates on a circular limit cycle. Here, the perturbations grow initially 
due to linear instability, then saturate nonlinearity due to $\Re(\zeta)>0$. The system develops a nonlinearly 
bounded oscillation with a stable, periodic limit cycle in $A$, the characteristic of a supercritical Hopf 
bifurcation. Each bifurcation diagram is illustrated and explained later. In compressible shear flow, this represents 
a Kelvin-Helmholtz vortex that self-saturates into a steady nonlinear wave. In Fig. \ref{fig-8b}, where $\Re(\zeta)<0$ 
and $\Im(\zeta)>0$ interprets a destabilizing nonlinearity and oscillatory behavior, exhibiting a subcritical Hopf 
bifurcation. The trajectory shows a rapid spiral outward with a blow-up. The negative nonlinear damping accelerates 
the amplitude growth. There is no saturation, so the wave amplitude explodes, modeling a subcritical instability or 
a nonlinearly unstable KHI. Physically, this implies the system transitions suddenly from equilibrium to unbounded 
growth, possibly indicating shock formation or transition to turbulence. In Fig. \ref{fig-8c}, where both $\Re(\zeta)<0$ 
and $\Im(\zeta)<0$ interprets blow-up with damping rotation and exhibits subcritical Hopf bifurcation. The trajectory shows 
the spiral outward and diverges. Here, both the growth and rotation are destabilized. The system shows spiral instability 
with no bounding, and it is an unstable node with a spiral structure consistent with subcritical blow-up behavior. This 
suggests an unbounded amplification of flow disturbances. In Fig. \ref{fig-8d}, where $\Re(\zeta)>0$ and $\Im(\zeta)<0$ 
interprets nonlinear damping and rotation in opposite directions, exhibits stable oscillations, and a supercritical Hopf 
bifurcation. The trajectory shows a spiral outward, which tends to saturate into a closed orbit. This is another 
supercritical Hopf bifurcation, but with the opposite sense of rotation due to negative $\Im(\zeta)$. The system shows 
oscillatory saturation in a nonlinear regime and represents a stable wave packet in the shear layer. In Fig. \ref{fig-8e}, 
where both $\Re(\zeta)>0$ and $\Im(\zeta)>0$, similar to (a) but with a smaller imaginary part, interpret slower rotation, 
exhibit a supercritical Hopf bifurcation. The trajectory shows a spiral outward, approaching a circular trajectory.  
A smoother transition to nonlinear saturation with a stable oscillatory amplitude can be interpreted here. Physically, 
represents marginal KHI that becomes self-regulated via nonlinear effects, indicating a stable, finite-amplitude wave. 
In Fig. \ref{fig-8f}, where $\Re(\zeta)<0$ and $\Im(\zeta)>0$ interpret negative damping, unstable spiral exhibits a 
subcritical Hopf bifurcation. The trajectory shows a spiral outward and accelerates growth. Here, the flow perturbations 
blow up without saturation, indicating subcritical bifurcation, likely near a turning point in a pitchfork-like structure. 
This corresponds to a transition to unstable secondary structures or turbulence in compressible shear flows. In Fig. 
\ref{fig-8g}, where $\Re(\zeta)>0$ and $\Im(\zeta)<0$ interpret stabilizing ensuring nonlinear saturation, exhibits a 
mild supercritical Hopf bifurcation. The trajectory shows a spiral tends to a circular orbit, representing a nonlinear 
wave packet that initially grows but saturates gracefully. Physically, its vortex roll-up in a slightly compressible 
environment that balances out over time. In Fig. \ref{fig-8h}, where both $\Re(\zeta)>0$ and $\Im(\zeta)>0$ interpret 
strong nonlinear saturation, exhibit a supercritical Hopf bifurcation. The trajectory shows a fast rotation due to 
large $\Im(\zeta)$. The flow is linearly unstable, but nonlinear damping begins quickly. The rotation suggests stable 
oscillations with physical implications in periodic wave shedding or coherent vortex patterns. In Fig. \ref{fig-8i}, 
where $\Re(\zeta)<0$ and $\Im(\zeta)>0$ interpret destabilizing nonlinearity due to negative $\Re(\zeta)$, and fast 
blow-up exhibits a subcritical Hopf bifurcation. The trajectory shows a divergent spiral. Physically, a catastrophic 
instability, possibly a post-bifurcation regime where the flow exhibits nonlinear breakdown, represents a breakup of 
coherent structures and the onset of turbulence.

%%%%%%%%%%%%%%%%%%%%%%%%%%%%%%%%%%%%%%%%%%%%%%%%%%%%%%
\begin{figure}
\BC
\subfigure[Supercritical Hopf]{\label{fig-9a}% 
\includegraphics[width=0.32\textwidth]{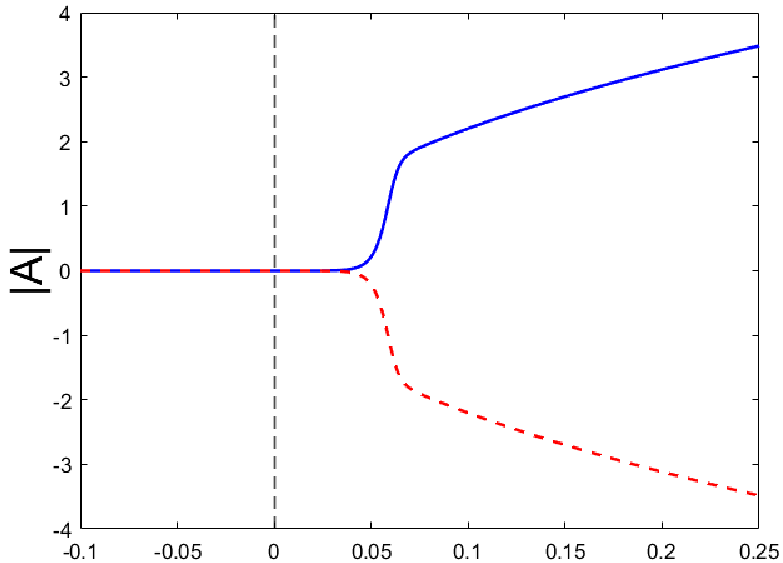}}\hfill%
\subfigure[Subcritical Hopf]{\label{fig-9b}% 
\includegraphics[width=0.3\textwidth]{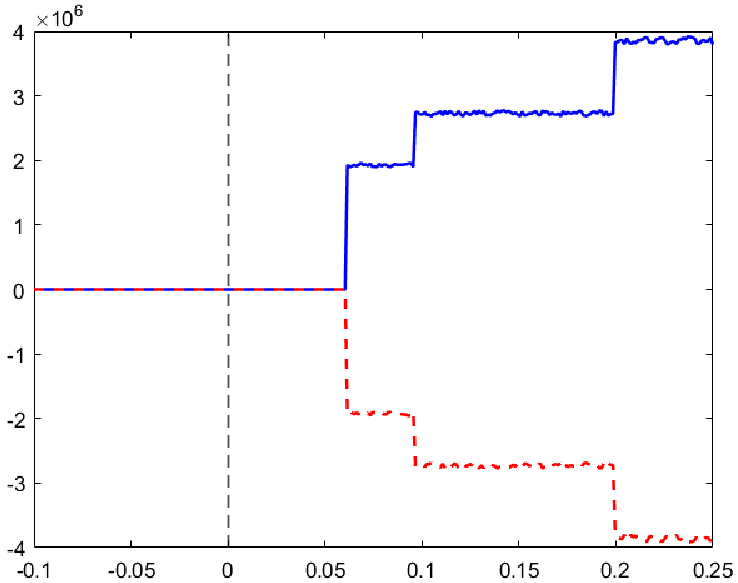}}\hfill%
\subfigure[Subcritical Hopf]{\label{fig-9c}% 
\includegraphics[width=0.3\textwidth]{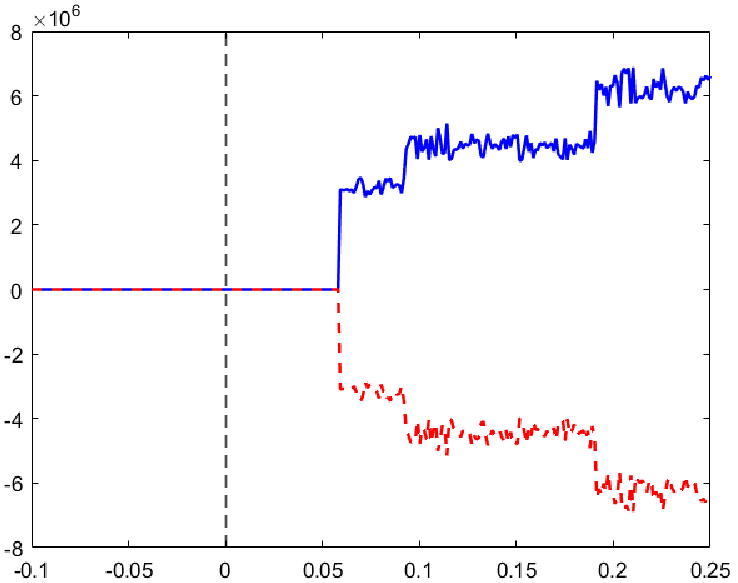}}\hfill%
\subfigure[Supercritical Hopf]{\label{fig-9d}% 
\includegraphics[width=0.32\textwidth]{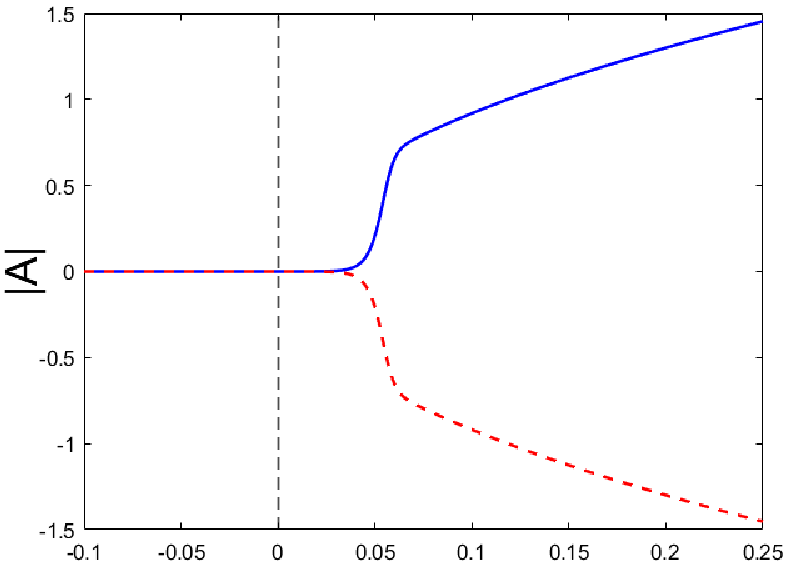}}\hfill%
\subfigure[Supercritical Hopf]{\label{fig-9e}% 
\includegraphics[width=0.3\textwidth]{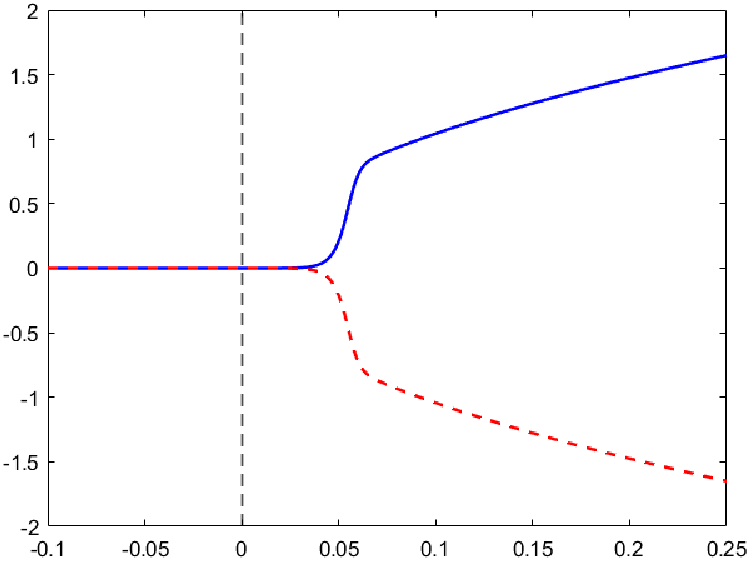}}\hfill%
\subfigure[Subcritical Hopf]{\label{fig-9f}% 
\includegraphics[width=0.3\textwidth]{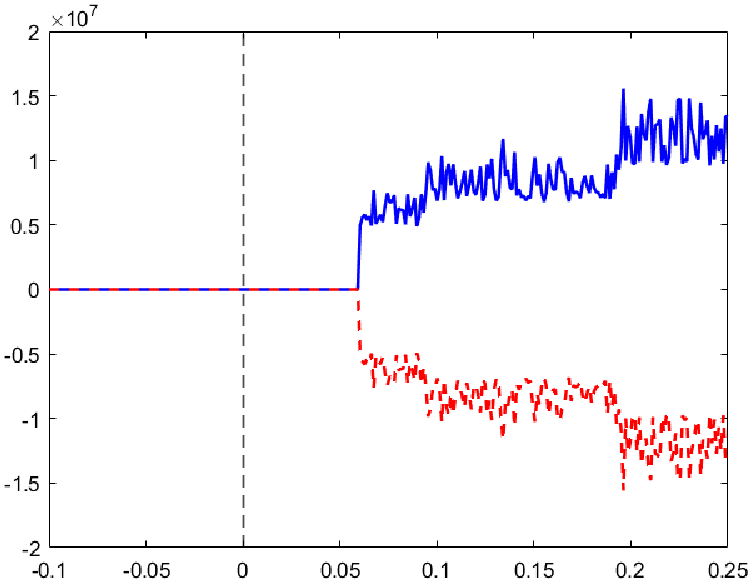}}\hfill%
\subfigure[Supercritical Hopf]{\label{fig-9g}% 
\includegraphics[width=0.32\textwidth]{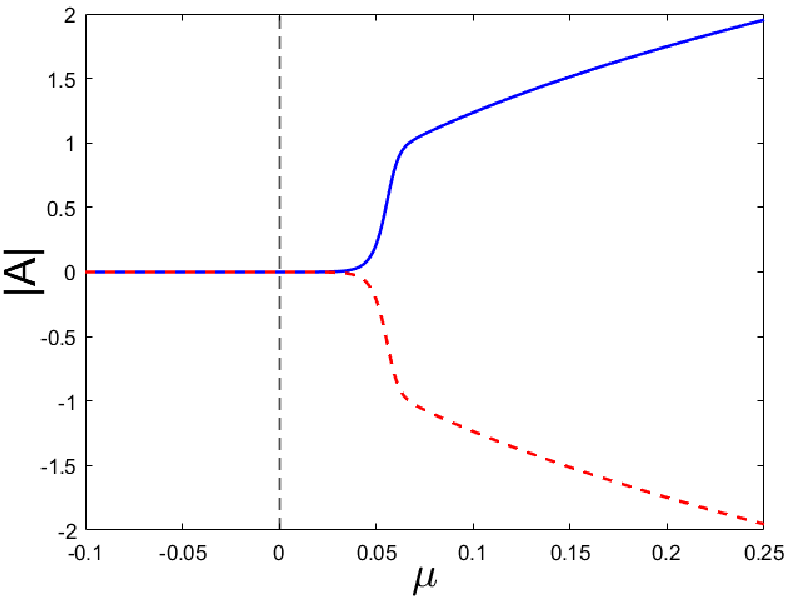}}\hfill%
\subfigure[Supercritical Hopf]{\label{fig-9h}% 
\includegraphics[width=0.3\textwidth]{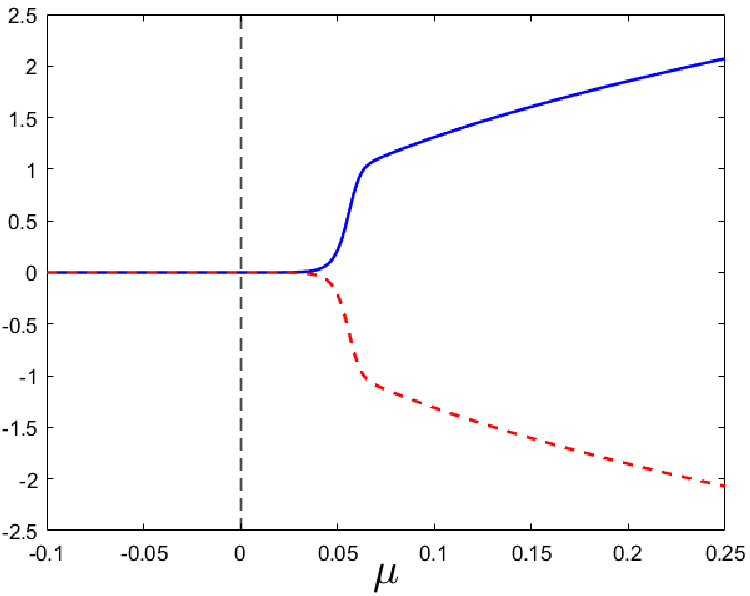}}\hfill%
\subfigure[Subcritical Hopf]{\label{fig-9i}% 
\includegraphics[width=0.3\textwidth]{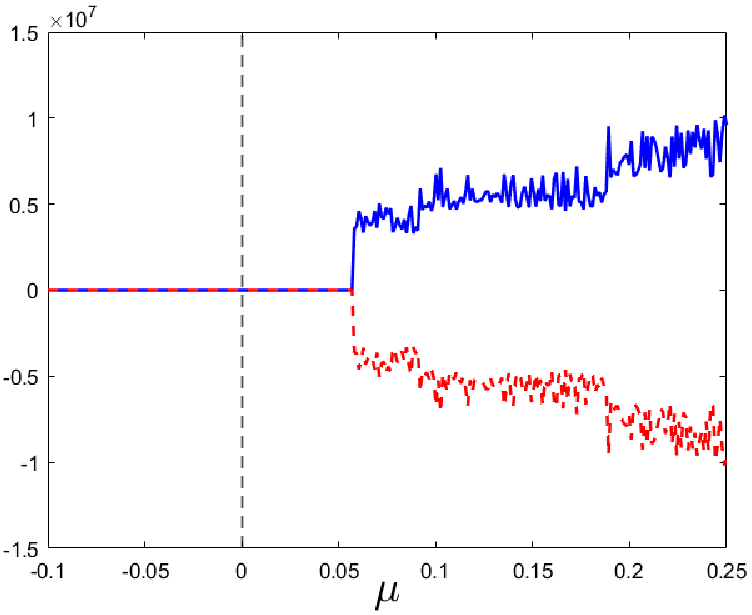}}
\caption{\label{fig:9}Subplots (a)-(i) show bifurcation diagrams illustrating the evolution of oscillation 
amplitude $|A|$ response as a function of the control parameter $\mu$, depending on the sign of $\Re(\zeta)$, 
for a compressible shear flow system governed by the Landau Eq. (\ref{amplitude}). Solid blue lines denote 
a stable oscillatory state, while dashed red lines represent unstable branches (limit cycles). The vertical 
dashed line at $\mu=0$ marks the linear instability threshold.}
\EC
\end{figure}
%%%%%%%%%%%%%%%%%%%%%%%%%%%%%%%%%%%%%%%%%%%%%%%%%%%%%%%%%
In Fig. \ref{fig:9}, bifurcation diagrams are shown for the amplitude response $|A|$ of flow disturbances 
evolving with respect to $\mu$ in compressible shear flows. Depending on the sign of the real part of $\zeta$, 
the Hopf bifurcation can be supercritical (smooth onset of stable oscillations) or subcritical (sudden jump to 
large amplitude oscillations with hysteresis or metastability). The estimated values of the Landau coefficient 
$\zeta$ and the linear growth $\mu$ for varying Mach numbers are presented in Table \ref{tab:zetavalues}. 
Panels (a), (d), (e), (g), and (h) exhibit the classical supercritical bifurcation shape with a smooth, 
continuous increase in oscillation amplitude beyond a critical point $\mu=0$. In Fig. \ref{fig-9a} (where 
$\Re(\zeta)>0$), as $\mu$ becomes positive, the flow transitions smoothly from a stable base state 
to a weakly nonlinear oscillatory regime. This reflects the early onset of instability due to weak compressibility. 
The oscillations saturate at a low amplitude, indicating strong damping from nonlinear effects. In Fig. 
\ref{fig-9d}, just beyond $M=1$, the bifurcation returns to a supercritical nature with a strongly positive 
$\Re(\zeta)$. This transition shows that the highly nonlinear instabilities near $M=1$ have relaxed 
into a more controlled regime. Perturbations grow and saturate smoothly, forming a stable limit cycle, 
indicating the onset of well-behaved compressible instabilities where pressure wave interactions are no 
longer disruptive but instead balance nonlinear growth. The flow regains predictability and resilience to 
external disturbances. In Fig. \ref{fig-9e}, the system shows a stable supercritical bifurcation 
with a moderate positive value of $\Re(\zeta)$. The nonlinear saturation is strong enough to limit 
the growth of oscillations in a smooth and gradual way. This regime suggests that the flow is robust to both 
linear and nonlinear perturbations, making it relatively easy to control. This flow regime is favorable for 
stable oscillatory behavior. The influence of compressibility is still present, but it does not lead to any 
irregular amplitude behavior or chaotic transitions. In Fig. \ref{fig-9g}, the bifurcation again turns 
supercritical at a slightly higher Mach number. The real part of $\zeta$ is positive, though smaller than at 
lower Mach numbers. This suggests modest nonlinear saturation, signifying oscillations grow predictably 
and stabilize over time. The amplitude remains bounded, and the system does not exhibit signs of bistability. 
Physically, this points to a self-regulating flow behavior in the high-subsonic to the low-supersonic regime, 
where instabilities grow but are controlled by inherent flow-damping mechanisms. In Fig. \ref{fig-9h}, the flow 
continues to exhibit supercritical behavior with steady amplitude saturation in this regime. The 
dynamics suggest a regime of regular, periodic oscillations that arise as soon as the linear instability 
begins. This behavior indicates a highly structured and relatively predictable flow, even in the high Mach 
number regime. Compressibility, while still present, seems to contribute to coherent structure formation 
rather than disorder. In compressible shear flow, this corresponds to stable and small amplitude oscillations 
that grow gradually as the control parameter Mach number increases.

Panels (b) exhibit the subcritical Hopf bifurcation, as shown by a negative real part of $\zeta$. This implies 
that finite-amplitude perturbations can trigger large, sustained oscillations even when the linear analysis suggests 
the base flow is stable (i.e., $\mu<0$). There is a jump in oscillation amplitude at some threshold $\mu>0$. 
Physically, this reflects a region of bistability and potential hysteresis where the flow can jump abruptly (to 
a large amplitude) from a quiescent state to a limit cycle or oscillatory state, depending on the amplitude of 
disturbances. Such behavior is typical in systems that show nonlinear effects that destabilize the flow early, and small 
disturbances may not trigger instability, but large ones will. 

Panels (c), (f), and (i) again correspond to subcritical bifurcation bits that show noisy or chaotic high amplitude 
states. In Fig. \ref{fig-9c}, at close to the threshold ($M\approx 1$), the flow exhibits a subcritical 
bifurcation but with more dramatic implications. The nonlinear coefficient $\zeta$ has a larger negative real 
part than in (b), indicating a stronger instability to finite perturbations. Physically, this suggests the presence 
of complex dynamics, possibly involving nonlinear resonance or even weak chaos. This condition amplifies 
compressibility effects, introducing feedback loops destabilizing the flow more aggressively. The oscillation 
amplitude might show irregular or noisy behavior, making control more challenging. In Fig. \ref{fig-9f}, at $M=1.3$, 
the system re-enters a subcritical regime, where nonlinear effects again destabilize the flow before linear stability 
sets in. The negative real part of $\zeta$ points to a situation where the flow might appear stable, but finite 
disturbances can trigger large amplitude oscillations. This hysteresis-prone regime is problematic for flow control, 
as it indicates that the system may jump to a highly oscillatory state with minimal external influence. This flow 
might support sudden bursts of energy or instability if slightly perturbed, resembling a nonlinear tipping point. 
In Fig. \ref{fig-9i}, the bifurcation is strongly subcritical at the highest Mach number considered, with the most 
negative real part of $\zeta$ in the dataset. This signifies a highly unstable nonlinear regime, where even small 
perturbations can lead to explosive amplitude growth. Physically, this may correspond to a flow on the verge of 
chaotic or turbulent transitions, where compressibility-driven feedback overwhelms the system's stabilizing 
mechanisms. It is likely to be highly sensitive to disturbances, and control strategies must be extremely 
robust to suppress or prevent catastrophic oscillations. These are scenarios where finite-amplitude disturbances 
create complex, sustained oscillations, even when the flow is theoretically linearly stable. These results interpret 
the alteration between supercritical and subcritical bifurcations with increasing Mach number, suggesting that 
compressibility has a non-monotonic influence on the nonlinear stability of the system. Certain regimes stabilize 
the flow through smooth saturation, while others enhance its sensitivity to perturbations and nonlinear effects. 

These nonlinear features are absent in the classical linear stability model. In contrast to the classical linear 
analysis by Blumen,\cite{Blumen70} which predicts a single-mode linear instability confined to a specific wavenumber--
Mach number regimes, our nonlinear analysis reveals the emergence of multiple instability branches and complex bifurcation 
behavior. Notably, the nonlinear framework captures secondary growth regimes and saturation dynamics entirely absent in 
Blumen's model. The presence of nonlinear self-interaction terms leads to amplitude modulation, multi-mode coupling, 
and saturation phenomena--including subcritical and supercritical bifurcation--that modify the growth rate and structure 
of perturbations. These findings highlight the crucial role of nonlinear effects in compressible supersonic flows and 
underscore the limitations of relying solely on linear theory for predicting long-term flow stability and transition.

\section{Conclusions\label{sec:conslusions}}
The onset and nonlinear evolution of instability in shear flows is classically described through the Kelvin-Helmholtz 
instability (KHI). This phenomenon becomes significantly richer in compressible flows due to the coupling between 
shear and compressibility. We analyzed the weakly nonlinear regime near the instability threshold, and we employed a 
Landau amplitude model derived from a multi-scale perturbation expansion. This analysis demonstrates how a Landau-Stuart 
equation captures the significant nonlinear dynamics of compressible KHI. The computed results include the identification 
of multiple bifurcation points with altering stability regimes, visual confirmation of bifurcation via phase portraits, 
Hopf bifurcation diagrams and quantitative and qualitative prediction of instability evolution using a low-order model 
are summarized below: 
\begin{itemize}
\item From the results of the dispersion relation, the figure shows that the dispersion relation essentially describes 
the perturbation of acoustic-like waves in a moving, compressible fluid with varying Mach numbers, and the figure 
visually captures the effect of flow direction and compressibility.
\item The existence of an optimal wavenumber for instability is a key feature, with the maximum growth rate shifting 
towards higher $k$ as $M$ increases. Our study shows that compressibility fundamentally modifies the instability 
characteristics, leading to a different neutral stability boundary and dominant mode selection.
\item The perturbed velocity and pressure results show how the compressibility effects fundamentally alter neutral 
stability curves. Our study demonstrates that compressibility can enhance instabilities, but wall effects dampen 
them (act as a stabilizing mechanism), balancing amplification and suppression. The present study more accurately 
models shear layers in confined environments, such as boundary layers over aircraft surfaces or compressible mixing 
layers in engineering flows. 
\item The reversal trend with increasing $M$ for the pressure distribution shows enhancement of pressure perturbations, 
implying that our imposed boundary conditions fundamentally alter how compressibility influences the flow. Instead 
of stabilizing the shear layer as in Blumen,\cite{Blumen70} compressibility intensifies perturbations, 
possibly leading to a different class of instability that is more relevant for high-speed aerodynamic applications, 
supersonic shear layers and turbulence modeling in confined flows. The observed trend suggests that compressibility 
effects alone can modify the nature of the instability, even in the absence of viscosity. 
\item The Landau coefficient $\zeta$ varies non-monotonically with Mach number, signifying alternating regions of 
supercritical and subcritical bifurcation. Combined with the oscillatory behavior of the linear growth rate $\mu$, 
this reveals a compressibility-mediated bifurcation structure with multiple critical points. These results demonstrate 
that compressibility introduces nontrivial modifications to both the linear onset and nonlinear evolution of the 
Kelvin-Helmholtz instability in shear flows. Time evolution of amplitude confirms the presence of both stable and 
unstable attractors, with nonlinear saturation highly sensitive to the Mach number.
\item The two- and three-dimensional phase portraits and trajectories across a range of Mach numbers reveal distinct 
dynamical regimes, including exponential growth, nonlinear saturation, decay, and sustained oscillations, each 
corresponding to variations in the real and imaginary parts of the Landau coefficient. As the Mach number increases, 
the system transitions through multiple bifurcation points, indicating alternating zones of stability and instability. 
Supercritical regimes, characterized by positive nonlinear damping ($\Re(\zeta)>0$), exhibit smooth amplitude 
saturation into stable, periodic vortex structures. In contrast, subcritical regimes ($\Re(\zeta)<0$) display 
unbounded growth, indicating nonlinear destabilization and potential transition to turbulence. The imaginary 
component of $\zeta$ induces phase rotation, indicating frequency modulation and oscillatory wave behavior. These 
results underscore the critical role of compressibility and nonlinearity in shaping the long-term dynamics and 
stability of shear-driven flows.
\item The bifurcation analysis reveals that compressible shear flows can undergo both supercritical and subcritical 
Hopf bifurcations, depending on system parameters. In the supercritical regime, oscillations emerge smoothly and 
predictably from the base state. In contrast, finite-amplitude instabilities arise abruptly in the subcritical regime, 
often accompanied by hysteresis or irregular large-amplitude behavior. These results highlight the critical role of 
nonlinear saturation mechanisms in dictating the transition dynamics, suggesting that linear stability analysis is 
insufficient and nonlinear stability analysis captures well the full spectrum of flow behaviors in compressible 
systems.
\end{itemize}
Overall, these findings highlight the critical role of Mach number in shaping the nonlinear dynamics and stability 
landscape of compressible shear layers. This study could improve predictive models for the boundary layer 
turbulence transition, where confined layers play a crucial role. It significantly modifies our understanding of 
instability growth, and modeling these dynamics is crucial for predicting flow behavior in practical applications, 
from jet flows to high-speed aerodynamics.

\begin{acknowledgments}
The authors would like to express their gratitude for the financial support from the Ministry of Science and 
Technology (MOST), Taipei, Taiwan (Grant No: 113-2811-M-001-105 and 113-2112-M-001-008). The authors acknowledge 
support for the CompAS Project from the Institute of Astronomy and Astrophysics, Academia Sinica (ASIAA), and 
for in-house access to high-performance computing (HPC) facilities. The authors also thank the National Center 
for High-performance Computing (NCHC) of National Applied Research Laboratories (NARLabs) in Taiwan for providing 
computational and storage resources.

\end{acknowledgments}

\section*{Declaration of Interests} 
The authors have no conflict of interest.

\section*{Data Availability Statement}
The data supporting this study's findings are available within the article.

\appendix
\section*{Appendixes}
\section{\label{appendix-2Sol} Solving the system of second-order perturbed equations and presenting the solutions}
The second-order perturbed system of equations is as follows:
\BA
\label{non-dim-2per1}
\frac{\P u_2}{\P t} + \bar{u}\frac{\P u_2}{\P x} + v_1\frac{\textrm{d}\bar{u}}{\textrm{d}y} + \frac{\P p_2}{\P x} &=& 
-\bigg(u_1\frac{\P u_1}{\P x} + v_1\frac{\P u_1}{\P y} = N_u\bigg)\,,\\
\label{non-dim-2per2}
\frac{\P v_2}{\P t} + \bar{u}\frac{\P v_2}{\P x} + v_1\frac{\P v_1}{\P y} + \frac{\P p_2}{\P y} &=& 
-\bigg(u_1\frac{\P v_1}{\P x} + v_1\frac{\P v_1}{\P y} = N_v\bigg)\,,\\
\label{non-dim-2per3}
M^2\bigg(\frac{\P p_2}{\P t} + \bar{u}\frac{\P p_2}{\P x}\bigg) + \frac{\P u_2}{\P x} + \frac{\P v_2}{\P y} &=& 
-\bigg[M^2\bigg(\frac{\P p_1}{\P t} + u_1\frac{\P p_1}{\P x} + v_1\frac{\P p_1}{\P y}\bigg) = N_p\bigg]\,,
\EA

We decompose the wave perturbation into normal modes for second-order terms, which is represented as
\BA
\label{normal-mode2}
u_2 = \hat{u}_2e^{2\i(k_xx + k_yy - \omega t)}\,, \qquad 
v_2 = \hat{v}_2e^{2\i(k_xx + k_yy - \omega t)}\,, \qquad 
p_2 = \hat{p}_2e^{2\i(k_xx + k_yy - \omega t)}\,,
\EA

Now, incorporating the solutions obtained from the first-order perturbed system of equations and the normal mode 
decomposition (\ref{normal-mode2}) into Eqs. (\ref{non-dim-2per1})-(\ref{non-dim-2per3}), the following algebraic 
system of equations yields,
\BA
\label{2nd-order-eq1}
(-2\i\omega + 2\i k_x\bar{u})\hat{u}_2 &=& -2\i k_x\hat{p}_2 + N_u\,,\\
\label{2nd-order-eq2}
(-2\i\omega + 2\i k_x\bar{u})\hat{v}_2 &=& -2\i k_y\hat{p}_2 - \hat{p}'_2 + N_v\,,\\
\label{2nd-order-eq3}
M^2(-2\i\omega + 2\i k_x\bar{u})\hat{p}_2 &=& -2\i k_x\hat{u}_2 - 2\i k_y\hat{v}_2 - \hat{v}'_2 + N_p\,.
\EA
Solving (\ref{2nd-order-eq1}) and (\ref{2nd-order-eq2}), we obtained the solutions for $\hat{u}_2$ and $\hat{v}_2$ as
\BA
\label{2Sol-1-2}
\hat{u}_2 = \frac{2k_x\hat{p}_2 + \i N_u}{2(\omega - k_x\bar{u})}\,, \qquad 
\hat{v}_2 = \frac{2k_y\hat{p}_2 + \i (N_v - \hat{p}'_2)}{2(\omega - k_x\bar{u})}\,,
\EA
Substituting (\ref{2Sol-1-2}) into (\ref{2nd-order-eq3}) will lead us to 
\BA
\label{2Sol-3eq}
\hat{p}''_2 + \bigg(4\i k_y + \frac{k_x\bar{u}'}{\omega - k_x\bar{u}}\bigg)\hat{p}'_2 
+ \bigg(4M^2\big(\omega - k_x\bar{u}\big)^2 - 4k^2 + \frac{2\i k_xk_y\bar{u}'}{(\omega - k_x\bar{u})}\bigg)\hat{p}_2 
&=& \bigg(2\i\big(k_xN_u + k_yN_v + (\omega - k_x\bar{u})N_p\big) 
+ \frac{k_xN_v\bar{u}'}{(\omega - k_x\bar{u})}\bigg)\,.\qquad
\EA
where $N_u$, $N_v$, and $N_p$ contains the quadratic nonlinear terms of $\hat{u}_1, \hat{v}_1, \hat{p}_1$ are as 
follows:
\BA
\label{nonlinear-expressions}
\NNM
N_u &=& \frac{k_x \hat{p_1}^2 \big(k_x k_y\bar{u}' + \i k^2 (\omega - k_x\bar{u}\big)}{(\omega - k_x\bar{u})^3}
+ \frac{k_x\hat{p_1}\hat{p_1}'\big(2k_y(\omega - k_x\bar{u}\big) - \i k_x\bar{u}'\big)}{(\omega - k_x\bar{u})^3}
- \frac{\i k_x\hat{p_1}'^2}{(\omega - k_x\bar{u})^2}\,,\\
\NNM
N_v &=& \frac{k_x^2\hat{p_1}\hat{p_1}'}{(\omega - k_x\bar{u})^2} + \frac{\i k_y k_x^2\hat{p_1}2}{(\omega - k_x\bar{u})^2}\,,\\
\NNM
N_p &=& \frac{\i k_x^2 M^2\hat{p_1}^2}{(\omega - k_x\bar{u})}\,.
\EA
We numerically solved Eq. (\ref{2Sol-3eq}), where for time and spatial derivatives, the Runga-Kutta Fourth-order 
method and a second-order central difference scheme are used.

\nocite{*}
\section*{References}
\bibliography{aipsamp}% Produces the bibliography via BibTeX.

\end{document}